\documentclass[journal]{IEEEtran}
\usepackage{amsmath,amsfonts}
\usepackage{algorithmic}
\usepackage{algorithm}
\usepackage{hyperref}
\usepackage{array,multirow}
\usepackage[caption=false,font=normalsize,labelfont=sf,textfont=sf]{subfig}
\usepackage{textcomp}
\usepackage{siunitx}
\usepackage{stfloats}
\usepackage{url}
\usepackage{booktabs}
\usepackage{verbatim}
\usepackage{graphicx}
\usepackage{cite}
\def\BibTeX{{\rm B\kern-.05em{\sc i\kern-.025em b}\kern-.08em
    T\kern-.1667em\lower.7ex\hbox{E}\kern-.125emX}}
\usepackage{balance}
\newcommand{\compressParag}{\looseness=-1}

\begin{document}
\title{An Unscented Kalman Filter-Informed Neural Network for Vehicle Sideslip Angle Estimation}
\author{Alberto Bertipaglia$^{1}$, Mohsen Alirezaei$^{2}$, Riender Happee$^{1}$ and Barys Shyrokau$^{1}$
\thanks{The Dutch Science Foundation NWO-TTW supports the research within the EVOLVE project (nr. 18484).}
\thanks{$^{1}$Alberto Bertipaglia, Riender Happee and Barys Shyrokau are with the Department of Cognitive Robotics, Delft University of Technology, 2628 CD Delft, The Netherlands
        {\tt\small A.Bertipaglia@tudelft.nl}, \tt\small {R.Happee@tudelft.nl} and \tt\small {B.Shyrokau@tudelft.nl}}%
\thanks{$^{2}$Mohsen Alirezaei is with the Department of Mechanical Engineering, University of Eindhoven, 5612 AZ Eindhoven, and with the Digital Industry-Software-Simulation and Testing Services, Helmond, The Netherlands {\tt\small m.alirezaei@tue.nl}}%
}

\markboth{}
{Shell \MakeLowercase{\textit{Bertipaglia et al.}}: An Unscented Kalman Filter-Informed Neural Network for Vehicle Sideslip Angle Estimation}

\maketitle

\begin{abstract}
This paper proposes a novel vehicle sideslip angle estimator, which uses the physical knowledge from an Unscented Kalman Filter (UKF) based on a non-linear single-track vehicle model to enhance the estimation accuracy of a Convolutional Neural Network (CNN). The model-based and data-driven approaches interact mutually, and both use the standard inertial measurement unit and the tyre forces measured by load sensing technology. CNN benefits from the UKF the capacity to leverage the laws of physics. Concurrently, the UKF uses the CNN outputs as sideslip angle pseudo-measurement and adaptive process noise parameters. The back-propagation through time algorithm is applied end-to-end to the CNN and the UKF to employ the mutualistic property. Using a large-scale experimental dataset of 216 manoeuvres containing a great diversity of vehicle behaviours, we demonstrate a significant improvement in the accuracy of the proposed architecture over the current state-of-art hybrid approach combined with model-based and data-driven techniques. In the case that a limited dataset is provided for the training phase, the proposed hybrid approach still guarantees estimation robustness.
\end{abstract}


\section{Introduction}
\IEEEPARstart{A}{ctive} vehicle control systems rely on the sideslip angle and yaw rate information to ensure stability and controllability \cite{chowdhri2021integrated, peterson2022exploiting}. Whereas low-cost gyro sensors measure the yaw rate, the vehicle sideslip angle must be estimated. Its direct measurement is possible via optical speed sensors or real-time kinematic positioning-global navigation satellite system (RTK-GNSS), but they are expensive to be installed in passenger vehicles \cite{liao2019adaptive}. Hence, the development of filter architectures is required to estimate the sideslip angle in real-time and with the desired accuracy error below one degree in high excitation driving conditions \cite{grip2009vehicle}.
Sideslip angle estimation is particularly challenging for the following aspects:
\begin{itemize}
    \item A large diversity of vehicle manoeuvres, e.g. steady-state, transient, low, and high excitation.
    \item The highly non-linear behaviour of tyres leads to a substantial limitation due to tyre model accuracy.
    \item Data collection requires expensive and high calibration sensitive instruments.
    \item Numerous external disturbances, e.g. bank angle, road slope, and road friction coefficient.
\end{itemize} 
\begin{figure}[!t]
    \centering
    \includegraphics[width=1\columnwidth]{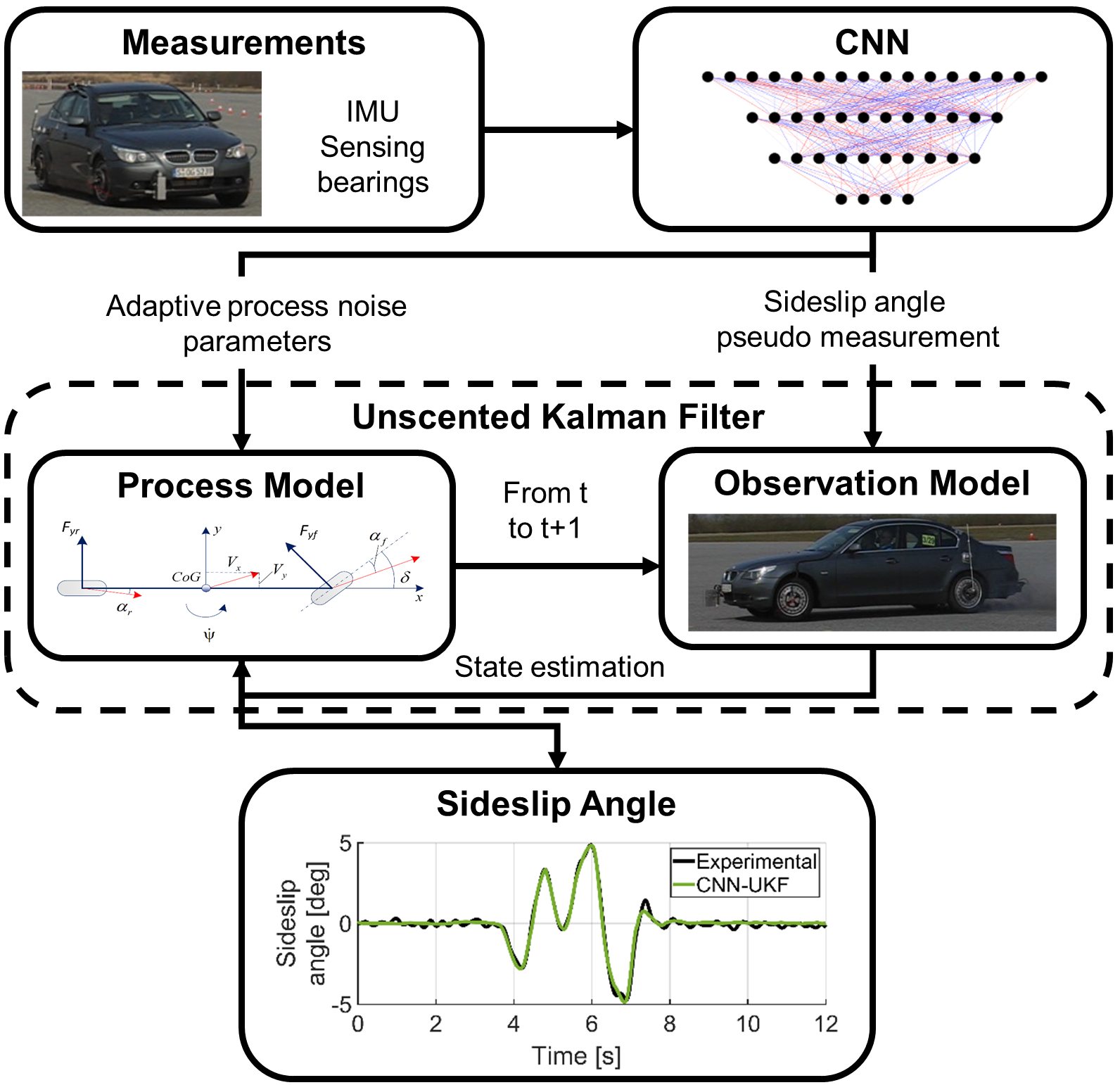}
    \caption{Framework overview of the CNN-UKF approach. A CNN provides a sideslip angle pseudo-measurement and the process noise parameters to a UKF based on a single-track vehicle model. The UKF monitors and weights the CNN's estimation through physical laws.}
    \label{fig:Framework}
\end{figure}

Several approaches have been proposed for vehicle sideslip angle estimation \cite{liu2020sideslip, viehweger2021vehicle}. 
They are split into three main groups, i.e. model-based, data-driven and hybrid approaches. 
The model-based approach relies on the physical knowledge of a vehicle model for state estimation. Open-loop deterministic models are insufficient to provide an accurate estimation, so stochastic closed-loop observers, e.g. extended Kalman filter (EKF), unscented Kalman filter (UKF), and particle filters, are currently applied to estimate unmeasurable states. EKF and UKF are the industrial state-of-art for vehicle sideslip angle estimation because their accuracy can be guaranteed in a specific operating region, and their properties are easily assessed \cite{graber2018hybrid}. However, they both struggle in transient and high excitation driving conditions due to the growing non-linearities in the vehicle model \cite{bertipaglia2022model}. Moreover, they require extensive system knowledge.
The data-driven approach has higher accuracy than the model-based approach when enough quality data are provided in the training phase \cite{bertipaglia2022model}. Different neural network (NN) architectures have been proposed, e.g. feed forward neural network (FFNN) and recurrent neural network (RNN). However, they all lack interpretability and generalisation capabilities. Thus, a purely data-driven approach is hardly applicable for safety applications in the automotive domain \cite{graber2018hybrid}.
The third approach, named hybrid, combines the pros of the model-based and data-driven approaches. It improves the model-based accuracy thanks to the NN outputs and, simultaneously, gives the data-driven approach an interpretability thanks to the vehicle model. In the proposed hybrid architectures\cite{kim2020vehicle, kim2021integrated, graber2018hybrid}, the model and the neural network work in a unidirectional way. Thus, the model in the hybrid architecture relies on the NN knowledge without backward communication, reducing the approach's potential.

This paper proposes a new hybrid approach for vehicle sideslip angle estimation. Its novelty is the mutualistic relationship between the model-based approach, characterised by a UKF based on a single-track model, and the data-driven approach, represented by a Convolutional Neural Network (CNN). It is a sequential hybrid architecture in which the CNN passes the pseudo-measurement of the sideslip angle, the level of distrust of its estimation and the process noise parameters' of the vehicle model to the UKF, see Fig. \ref{fig:Framework}. A key aspect of the proposed hybrid approach is the training process which allows the development of a physics-informed NN \cite{raissi2019physics}. This paper contains the non-linear vehicle dynamics in a UKF, so the physics-informed NN will also be referenced as a UKF-informed NN. The training is end-to-end, so the Back-Propagation Through Time (BPTT) algorithm moves through the CNN, the UKF and backwards. Thus, the CNN is constrained to respect the physical laws of vehicle dynamics. Furthermore, it allows the CNN to estimate variables for which a reference is unavailable, i.e. the process noise parameters and pseudo-measurement level of distrust. This will lead to a high estimation accuracy compared to the state-of-art hybrid approaches. The performance is evaluated using a large-scale real-world experimental dataset. The dataset contains a great diversity of driving situations, recorded with a constant high friction coefficient.\newline

The paper is organised as follows. Section \ref{pw} contains a summary of the previous works and the main paper contributions. Section \ref{CNNUKF} describes the CNN and UKF used in the proposed hybrid approach. Section \ref{exp} describes how the experiments are conducted and evaluated. Section \ref{res} reports the obtained results, and Section \ref{conc} summarises the conclusions and future research paths.

\section{Related Works}
\label{pw}
\begin{table}[!t]
    \label{tab:Overview}
    \caption{Overview of the vehicle sideslip angle estimation approaches.}
    \begin{center}
    \begin{tabular}{ >{\centering\arraybackslash}m{0.5in}  >{\centering\arraybackslash}m{2.7in}}
    \toprule[1pt]
    \textbf{Approach} & \textbf{Features \& Authors}\\
    \hline
    \multirow{6}{*}{Model-based} & Kalman filter based on a kinematic model \cite{tuononen2009vehicle,madhusudhanan2016vehicle,selmanaj2017vehicle} \\
    
    & Kalman filter based on IMU \& GNSS measurements \cite{liu2021,xia2021autonomous,li2021event,ding2021event,wischnewski2019vehicle,park2022vehicle, song2021vehicle, yoon2013robust} \\
    
    & EKF based on a dynamic model \cite{van2018adaptive,gadola2014development,vaseur2021robust,li2014variable,cheli2015smart,van2017vehicle,ando2022localization}\\
    
    & UKF based on a dynamic model \cite{doumiati2010onboard,antonov2011unscented,heidfeld2021optimization,singh2019vehicle,bechtoff2016cornering,mazzilli2021benefit} \\
    
    & Hybrid based on dynamic \& kinematic models \cite{liao2019adaptive,galluppi2018mixed,villano2021cross} \\
    
    & Online gradient descent \cite{chen2018vehicle} \\
    
    & Modular estimation scheme \cite{hashemi2017corner, jalali2017integrated} \\
    \hline
    \multirow{3}{*}{Data-driven} & FFNN \cite{sasaki2000side,melzi2011vehicle}\\
    
    & RNN \cite{bonfitto2020combined,ghosh2018deep,liu2020time}\\
    
    & ANFIS \cite{boada2015sideslip}\\
    
    & Kernel-based LPV \cite{breschi2020vehicle}\\
    \hline
    \multirow{9}{*}{Hybrid} & FFNN, ANFIS \& UKF \cite{acosta2017robust, novi2019integrated, boada2016vehicle} \\
    
    & RNN (GRU) \& UKF \cite{sieberg2019hybrid} \\
    
    & Differentiable EKF \cite{piga2021differentiable}\\
    
    & Kalman filter \& FFNN \cite{jeon2021simultaneous}\\
    
    & Piecewise Affine \& Takagy-Sugeno \cite{zhang2021novel}\\
    
    & FFNN \& Kalman in the back-propagation \cite{haarnoja2016backprop}  \\
    
    & KalmanNet \cite{revach2022kalmannet} \\
    
    & Kinematic model \& RNN (GRU) \cite{graber2018hybrid} \\
    
    & Deep Ensemble Network (LSTM) \& EKF \cite{kim2020vehicle}, UKF \cite{kim2021integrated} \\

    \bottomrule[1pt]
    \end{tabular}
    \end{center}
\end{table}
\noindent A summary of the three categories, i.e. \emph{model-based}, \emph{data-driven} and \emph{hybrid}, is presented in Table \ref{tab:Overview}.

The first approach is called \emph{model-based} and relies on the laws of physics. The vehicle behaviour can be described using the geometric constraints, i.e. kinematic model, or considering the forces and moments acting on the vehicle, i.e. dynamic model. The kinematic model requires only geometrical parameters and does not need extensive vehicle parametrisation because its reliability depends mainly on sensing capabilities. The state-of-art kinematic observer \cite{selmanaj2017vehicle} is based on a linear parameter varying system, where the states are the vehicle velocities, and the accelerations are the inputs. This approach leads to high accuracy in transient manoeuvres, but the model is not observable in nearly steady-state conditions \cite{liao2019adaptive}. Hence to avoid unobservability, a heuristic function is applied to lead the lateral velocity to zero when the vehicle is moving straight or nearly straight \cite{selmanaj2017vehicle}. The downside is the amount of data necessary to define the heuristic function. Moreover, despite the performance improvement, it is still susceptible to integration error and sensor drifting. Thus, in recent publications \cite{liu2021,xia2021autonomous,li2021event,ding2021event}, the measurements from the Inertial Measurement Unit (IMU) are coupled with those from a Global Navigation Satellite System (GNSS) to increase the amount of information available for the estimator. The velocities measured by the GNSS are integrated into an estimation-prediction framework, which estimates the sideslip angle and partially compensates for the error induced by the low GNSS sampling rate \cite{liu2021}. However, GNSS/IMU fusion kinematic approach still suffers from the low GNSS sampling rate. Furthermore, a high-precision GNSS is too expensive as the standard sensor in passenger vehicles, and signal reception cannot always be assured. Therefore, it is mainly applied to racing \cite{wischnewski2019vehicle}.
Thus, a solution is to consider dynamic models to rely less on the sensor signal quality. Dynamic models allow a more robust noise computation of the accelerations than kinematic models \cite{liao2019adaptive}. However, dynamic models require a more profound knowledge of vehicle parameters and the presence of a tyre model, which is a critical source of uncertainty \cite{mazzilli2021benefit}. 
EKF and UKF are the state-of-art estimation techniques for the model-based approach, and the process and the observation noises are commonly assumed to be Gaussian and uncorrelated. The EKF uses a first-order Taylor series expansion to linearise around the current mean and covariance. It has excellent accuracy in nearly steady-state conditions and when the vehicle behaves close to linearity, i.e. up to a lateral acceleration of approximately \SI{4}{m/s^2} \cite{bertipaglia2022model}. When the vehicle behaves with strong non-linearities, UKF assures a better estimation accuracy because it linearises up to the third order of the Taylor series expansion \cite{bertipaglia2022model}. However, both observers suffer from the mismatches between the physical and modelled tyre behaviour.
A possible solution is to combine the pros of dynamic and kinematic models to develop a hybrid kinematic-dynamic observer \cite{galluppi2018mixed, villano2021cross}. This family combines the accuracy in transient manoeuvres of the kinematic models and the better robustness to sensor noise of the dynamic models. The kinematic and the dynamic filters work simultaneously, and the final sideslip angle estimation is a weighted average of the two approaches. The weights are chosen according to the lateral acceleration signal \cite{villano2021cross}. However, the weighting coefficients' tuning is complex, and the optimum solution varies according to the considered manoeuvres.
Another solution to combine dynamic and kinematic models is the development of a modular scheme to estimate in sequential steps tyre forces, longitudinal and lateral velocities \cite{hashemi2017corner}. It consists of monitoring the wheel capacities under combined slip at each vehicle corner to estimate the individual forces and velocities. The approach is experimentally validated in different road conditions, but the results do not show its performance when the vehicle is driven at the limit of handling. Thus, the approach applicability to evasive manoeuvres is limited.

A solution to enhance the state estimation robustness to tyre model inaccuracies of dynamic model is the introduction of adaptive tyre models \cite{van2018adaptive, acosta2017robust} or new proprioceptive load-sensing technology, e.g. intelligent bearings or smart tyres \cite{kerst2019model, mazzilli2021benefit}. The Kalman filter can use tyre force measurements as an additional feedback to improve the estimation and magnify the Kalman gain, especially in the case of non-linear vehicle behaviour. The enhanced vehicle safety and the sensor's cost efficiency make the innovative load-sensing technologies candidate to become part of the standard sensor setup for passenger vehicles \cite{mazzilli2021benefit}.

A \emph{data-driven} approach reduces extensive requirements of system knowledge compared to the model-based approach. A deep NN with eight hidden layers, each having a different number of long short-term memory (LSTM) cells, is proposed \cite{ghosh2018deep}. Despite the increased training time of such a deep NN, the authors state that a smaller network was incapable of reaching the level of accuracy of deeper NN. The issue is that deep NNs are prone to overfit, and their performance strongly lacks generalisation capabilities. To overtake this issue, an NN classification is applied to choose which available NN is most suitable for specific road conditions \cite{bonfitto2020combined}. Each of the three FFNNs is built with a single hidden layer, and they are trained with three different datasets corresponding to three different road friction conditions, i.e. dry, wet and icy. Moreover, the performance of data-driven approach can be enhanced by the availability of tyre force measurements \cite{bertipaglia2022model}. In this case, a FFNN with two hidden layers outperforms the accuracy of a more complex RNN architecture based on LSTM cells. A FFNN also exceeds the performance of various model-based approaches, even if it tends to sporadic higher maximum error. However, the data-driven performance is highly dependent on the amount of representative data, and the data-driven approach will lack performance as soon as the dataset contains a lower amount of data in a particular range of the sideslip angle.

Although the data-driven approach generally has a better estimation performance than the model-based approach, it is impossible to guarantee robust performance over vehicle operating conditions. Conversely, a model-based approach based on a dynamic model with tyre force measurements has lower accuracy, but its performance is consistent over the working region \cite{bertipaglia2022model}. Thus, a \emph{hybrid} model-based and data-driven approach is proposed. Here we employ two leading typologies: model-to-NN and NN-to-model as explained below.

Model-to-NN family aims to augment the number of the NN's inputs using the output of a vehicle model. This will transfer some immeasurable physical states to a NN. A kinematic vehicle model can compute the derivative of the sideslip angle, which is used as extra input for the following RNN based on a Gated Recurrent Unit (GRU) cell \cite{graber2018hybrid}. The kinematic model provides the NN with a pseudo-measurement that contains a lot of errors, biases and drift. Despite this, with the extra vehicle model information, the NN reduces the sideslip angle's Mean Squared Error (MSE) of the non-informed NN by \SI{2.7}{\%}, \SI{5.6}{\%}, and \SI{1.2}{\%}, respectively, for dry, wet and snow conditions \cite{graber2018hybrid}. The slight improvement shows the benefits of developing a hybrid approach and highlights the importance of providing a more accurate pseudo-measurement.

NN-to-model family, vice-versa, aims to provide a sideslip angle pseudo-measurement to the following EKF/UKF. In this case, the NN output is post-processed by a Kalman filter to improve the sideslip angle estimation. One of the first approaches combines an Adaptive Neuro-Fuzzy Inference System (ANFIS) \cite{boada2015sideslip} with a UKF to estimate sideslip angle. The model-based component is employed as a filter to minimise the noise of the NN output and the variance of the estimation mean square error \cite{boada2016vehicle}. The ANFIS is trained using synthetic data generated through a high-fidelity simulation environment (CarSim). The ANFIS-UKF improves the performance of only an ANFIS \cite{boada2015sideslip} by \SI{21}{\%} on average for five manoeuvres with high friction conditions. However, the presented figures show a maximum value of sideslip angle of only \SI{3}{deg} in absolute value, which makes the estimation performance easier than in extreme driving conditions.
Furthermore, there is no explanation for tuning the observation noise parameter related to the pseudo-sideslip angle. This value is essential because it defines the level of distrust the UKF can give to the NN output. A similar approach involves integrating an FFNN with a UKF based on a kinematic vehicle model \cite{novi2019integrated}. Contrary to the ANFIS-UKF approach, the model-based component of the hybrid approach is responsible for filtering the estimation noise and correcting the NN output. This is possible thanks to a proportional feedback correction which improves the performance of the pseudo-sideslip angle. Although the NN is trained using synthetic measurements, the approach is validated using experimental data and the presented results. Unfortunately, all the results are normalised, so it is impossible to understand if the vehicle was driven at the limit of handling. The presented approach improves the data-driven approach estimation accuracy of \SI{73.3}{\%}. However, there needs to be an analysis of how to decide the distrust level of the pseudo-sideslip angle. Otherwise, when the NN is uncertain due to a lack of data, it will negatively influence the UKF's performance. Furthermore, the kinematic vehicle model is highly susceptible to measurement noise and is not observable in steady-state driving. For this reason, the FFNN is substituted by a deep ensemble NN in more recent publications \cite{kim2020vehicle, kim2021integrated}.

Deep Ensemble (DE) of RNN, based on LSTM cells, estimates a sideslip angle pseudo-measurement and its level of distrust which are then provided as extra measurements to a UKF \cite{kim2020vehicle, kim2021integrated}. The level of distrust is modified through a user-defined linear function before being used by the UKF. This further step is mandatory to scale the NN's distrust level to a meaningful value for the UKF. This hybrid architecture reduces the Root Mean Squared Error (RMSE) by, on average, \SI{8}{\%} vs the RNNs\cite{kim2020vehicle}. The extra tuning of the level of distrust can easily lead the approach to overfit.
Moreover, the level of distrust is computed through the standard deviation of the sideslip angle pseudo-measurements estimated by the RNNs. This does not lead to a physics-informed NN, so it is still complex to assess the properties of this hybrid approach. The reason is that the DE-RNN is not aware of the performance of the UKF, so the estimated level of distrust is not scaled according to the UKF's accuracy. Vice-versa, a physics-informed NN learns the Kalman filter's precision during the training, providing the best level of distrust to maximise the hybrid sideslip angle estimation.

This work proposes a hybrid approach employing a mutual relationship between the model-based and data-driven approaches for vehicle sideslip angle estimation. The inputs and outputs of the NN are, respectively, inputs and measurements of the UKF. The mutual relationship is enforced by the end-to-end training meaning that the back-propagation algorithm passes through the NN, UKF and vice-versa.

The main contribution of this paper is threefold\footnote{The code for our method will be released upon paper acceptance.}. The first is a mutual hybrid approach in which the CNN, trained end-to-end with the UKF, estimates a pseudo-measurement of the sideslip angle and its level of distrust. The UKF observation model uses both NN outputs to enhance the accuracy (MSE) of the state-of-art model-based, data-driven, and hybrid approach for vehicle sideslip angle estimation. Contrary to other hybrid approaches \cite{kim2020vehicle,kim2021integrated}, our NN's level of distrust does not need to be scaled after the NN training due to the mutualistic relationship of the architecture.

The second contribution is the online estimation of the UKF process model uncertainties due to the mutualistic hybrid approach. This provides higher robustness than the state-of-art, even when a limited dataset is used for the training.

The third contribution is that the proposed hybrid architecture is a UKF-informed NN, which means that the NN must comply with the laws of physics of vehicle dynamics. Thus, the proposed approach has a lower Maximum Error (ME) than a state-of-art model-based \cite{mazzilli2021benefit}, and data-driven \cite{bertipaglia2022model} approaches as well as the state-of-art hybrid approach \cite{kim2020vehicle}.

\section{UKF-Informed Neural Network}
\label{CNNUKF}
\begin{figure}[!t]
    \centering
    \subfloat[\label{fig:UKF_CNN}]{
        \includegraphics[height=2.8cm, keepaspectratio]{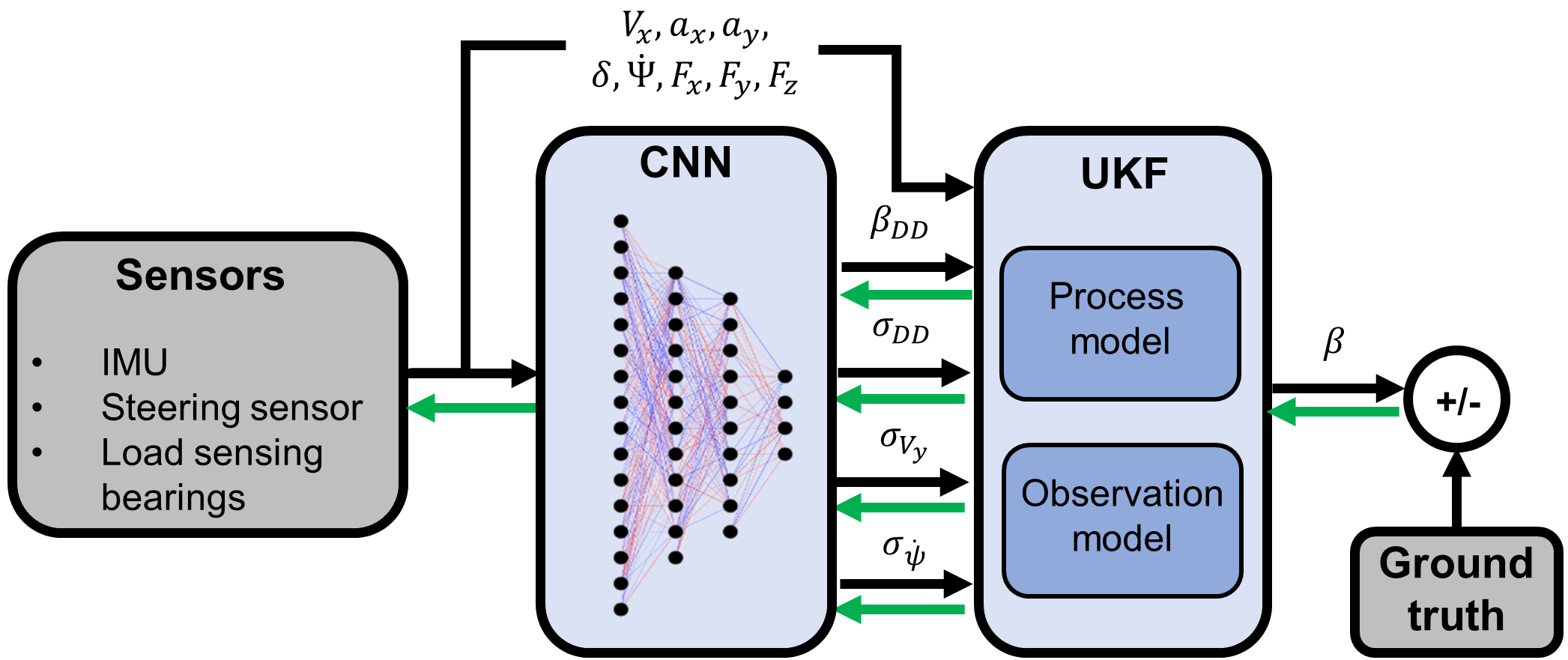}} \\
    \centering
    \subfloat[\label{fig:Kim}]{%
       \includegraphics[width=8.8cm, keepaspectratio]{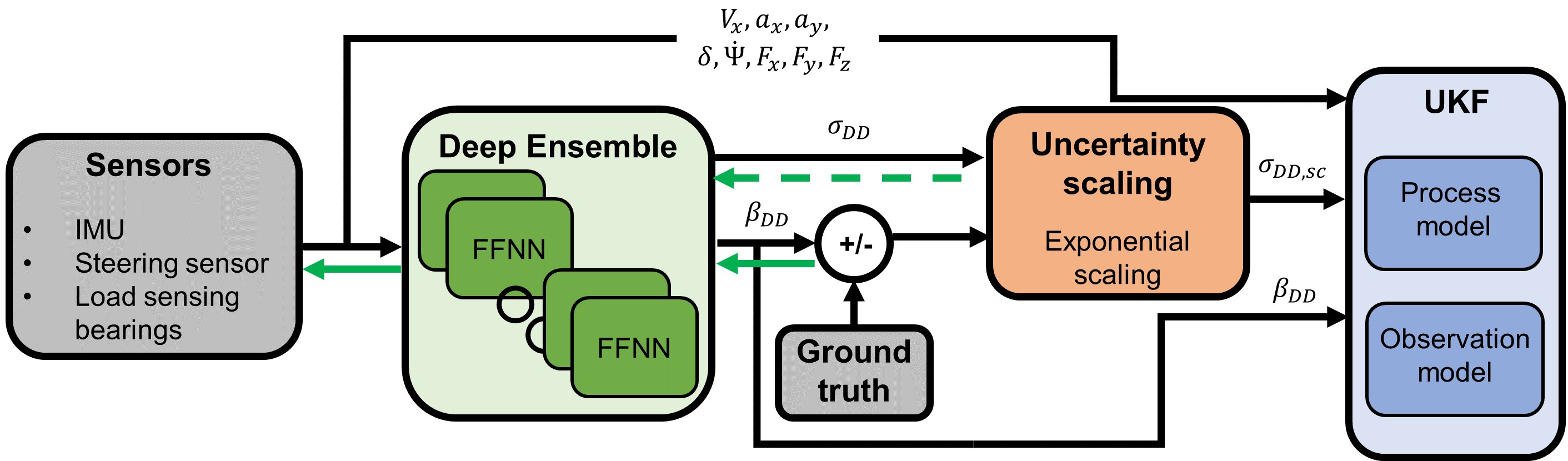}}
       \caption{Fig. \ref{fig:UKF_CNN} shows the proposed hybrid approach architecture (CNN-UKF). Fig. \ref{fig:Kim} shows the baseline hybrid architecture (DE-UKF) proposed by \cite{kim2020vehicle}. The black arrows show the flow of information during the online estimation, while the green arrows show the flow of information during the back-propagation. The dashed green arrow represents a term used in the cost function computation but not used by the optimiser to update the NN weights.}
    \label{fig:Comparison}
\end{figure}

\noindent This section describes the proposed hybrid approach based on a CNN end-to-end trained with a UKF (CNN-UKF). A comparison between the proposed mutualistic hybrid approach and the hybrid unidirectional baseline \cite{kim2020vehicle} is represented in Fig. \ref{fig:UKF_CNN} and Fig. \ref{fig:Kim}, respectively. The proposed approach develops a UKF-informed NN, where the NN is constrained to respect the vehicle dynamics. At the same time, the baseline (DE-UKF) corresponds to a UKF augmented by the DE outputs.
The approach's discretisation is performed through a zero-order hold method \cite{hashemi2017corner} due to its good trade-off between simplicity and accuracy. The discretisation works at \SI{100}{Hz}, the standard frequency for vehicle state estimation.
\subsection{Data-driven Component}

\noindent A straightforward CNN can cope with the complexity of the task because the approach's strength is inside the hybrid architecture. It consists of an input layer, two hidden layers and an output layer.

Seventeen measurements form the input layer $\left(x\right)$: longitudinal and lateral accelerations $a_x$ and $a_y$ respectively, longitudinal velocity $V_x$, road wheel angle $\delta$, yaw rate $\dot\psi$, and longitudinal, lateral and vertical tyre forces for each of the four wheels, respectively $F_x$, $F_y$ and $F_z$. Before being used, the input measurements are normalised because each input has a different physical meaning and order of magnitude. Thus, all the inputs are mapped onto the interval $[0,\,1]$ to speed up and stabilise the training process \cite{brownlee2018better}. A different normalisation method which scales the data to a mean of zero and a standard deviation of one has been tested. Still, the mapping onto the interval $[0,\,1]$ produced the best results after the training.

The two hidden layers consist of 200 and 100 neurons and Rectified Linear Unit (ReLU) activation functions. The hidden layers are 2D convolutions with kernel sizes $1\times1$, $0$ padding, stride equal to $1$ and active bias. The CNN uses a dropout regularisation technique equal to $0.2$ and a Xavier weight initialisation to avoid overfitting. 

The output layer is formed by four neurons corresponding to the pseudo-measurement of the sideslip angle $\beta_{DD}$, the level of distrust in the pseudo-measurement $\sigma_{DD}$, the uncertainty of the UKF process model lateral velocity $\sigma_{V_y}$ and the uncertainty of the UKF process model $\sigma_{\dot\psi}$. A reference is available only for $\beta_{DD}$, but the other three outputs strongly affect the estimation of the model-based component, which is used in the training loss function; see Section \ref{training} for further details. Thus, all four CNN outputs are correctly trained during the end-to-end training. $\sigma_{V_y}$, $\sigma_{\dot\psi}$ and $\sigma_{DD}$ are further processed with a sigmoid function to constrain their values inside the meaningful interval $[0,\,1]$. This last step assures that the CNN can produce uncertainties which not lead to UKF failure.

\subsection{Model-based Component}
\begin{figure}[!t]
    \centering
    \includegraphics[height=3cm, keepaspectratio]{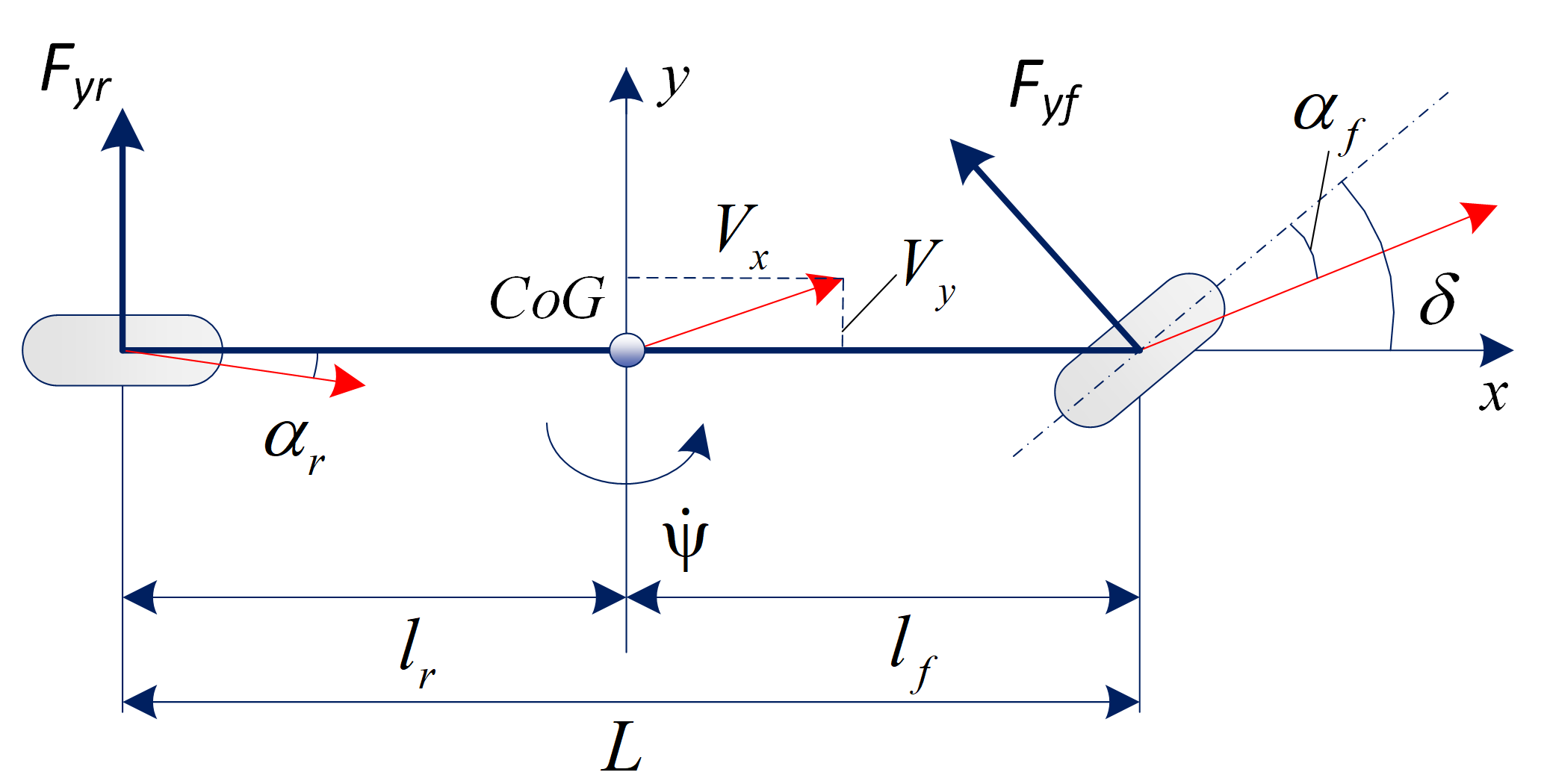}
    \caption{Single-track vehicle model.}
    \label{fig:bicycle}
\end{figure}
\noindent A UKF based on a non-linear single-track model with tyre axle forces computed by the Dugoff tyre is chosen as the model component of this study \cite{bertipaglia2022model}, see Fig. \ref{fig:bicycle}. The Dugoff tyre model parameters (tyre cornering stiffness, peak friction coefficient and velocity reduction friction coefficient) are optimised offline using experimental skidpad measurements\cite{doumiati2010onboard, mazzilli2021benefit, heidfeld2021optimization}. The implemented optimisation is a genetic algorithm due to its efficiency with a non-linear and non-convex cost function. The vehicle's symmetry is exploited to merge the left and right wheels into a single central axle, which emulates the entire vehicle's behaviour. This model considers only the in-plane dynamics, so the lateral weight transfer, roll and pitch dynamics are ignored. The static weight distribution is considered together with the effect of steady-state longitudinal weight transfer concerning the normal forces on the front and rear axle. A UKF is implemented for its superior estimation accuracy when the vehicle behaves strongly non-linearly. The vehicle states ($x_s$) are the $V_y$ and the $\dot{\psi}$, while the vehicle inputs ($u_v$) are the $V_x$ and the $\delta$. The stochastic process model is responsible for predicting the next time steps of the states according to the following equation:
\begin{equation}
    \dot{x}_s\left(t\right) = f\left(x_s\left(t\right), u_v\left(t\right)\right) + \omega\left(t\right)
    \label{eq_pro}
\end{equation}
where $f\left(x_s\left(t\right),u_v\left(t\right)\right)$ is the non-linear single track vehicle model, eq. \ref{eq_ve}, and $\omega$ is the vector containing the process noise parameters $[\sigma_{V_{y}},\,\sigma_{\dot\psi}]$.
\begin{equation}
    \begin{split}
    &f\left(x_s, u_v\right) = \\
    &=\begin{cases}
        \dot{V}_y = \frac{1}{m} \left(F_{yr}\left(x_s,u_v\right) \cos\left(\delta\right) + F_{yr}\left(x_s,u_v\right)\right) - V_x\dot{\psi}\\
        \ddot{\psi} = \frac{1}{I_{zz}} \left(l_f F_{yf}\left(x_s,u_v\right) \cos\left(\delta\right) - l_r F_{yr}\left(x_s,u_v\right)\right) \\
    \end{cases}
    \end{split}
    \label{eq_ve}
\end{equation}
where $m$ (\SI{1970}{kg}) is the vehicle mass, $I_{zz}$ (\SI{3498}{kg m^2}) is the vehicle moment of inertia about the vertical axis, $l_f$ (\SI{1.47}{m}) and $l_r$ (\SI{1.41}{m}) are, respectively, the distance of front and rear axles from the vehicle CoG. $F_{yf}$ and $F_{yr}$ are, respectively, the lateral tyre forces at the front and rear axles. The process noise parameters, $\sigma_{V_{y}}$ and $\sigma_{\dot\psi}$, are assumed Gaussian and uncorrelated and they capture the uncertainties due to:
\begin{itemize}
    \item The mismatch between the physical and modelled vehicle behaviour.
    \item The discretisation error.
    \item The various operational environments in which the sensors operate.
\end{itemize} 
The filter performance is strongly connected with the process noise parameters, so these are initially tuned using a two-stage Bayesian optimisation (TSBO) \cite{bertipaglia2022two}. During the estimation, they are computed online by CNN. This is only possible thanks to the mutualistic relationship between CNN and the UKF.

The observation model is responsible for comparing the process model predictions with the available measurements, according to the following equation.
\begin{equation}
    y_m\left(t\right) = g\left(x_s\left(t\right), u_v\left(t\right)\right) + v\left(t\right)
    \label{eq_ob}
\end{equation}
where $g\left(x_s\left(t\right),u_v\left(t\right)\right)$ is the measurement vehicle model, eq. \ref{eq2}, and $v$ is the vector containing the observation noise parameters $[\sigma_{a_{y\,me}},\,\sigma_{\dot{\psi}_{me}},\,\sigma_{F_{yf\,me}},\,\sigma_{F_{yr\,me}},\,\sigma_{DD}]$.
\begin{equation}
    \begin{split}
    &g\left(x_s, u_v\right) =\\ 
    &=\begin{cases}
        a_{y\,me} = \frac{1}{m} \left(F_{yf}\left(x_s,u_v\right) \cos\left(\delta\right) + F_{yr}\left(x_s,u_v\right)\right) \\
        \dot{\psi}_{me} = \dot{\psi}\\
        F_{yf\,me} = F_{yf}\left(x_s,u_v\right) \\
        F_{yr\,me} = F_{yr}\left(x_s,u_v\right) \\
        \beta_{DD} = \textnormal{atan}(\frac{V_y}{V_x})\\
    \end{cases}
    \end{split}
    \label{eq2}
\end{equation}
where $a_{y\,me}$, $\dot{\psi}_{me}$, $F_{yf\,me}$ and $F_{yr\,me}$ are the vehicle measurements, and $\beta_{DD}$ is the pseudo-measurement, corresponding to the CNN's output. The observation noise parameters $\sigma_{a_{y\,me}}$ (\SI{0.033}{m/s^2}), $\sigma_{\dot{\psi}_{me}}$ (\SI{0.001}{rad/s}), $\sigma_{F_{yf\,me}}$ (\SI{26}{N}) and $\sigma_{F_{yr\,me}}$ (\SI{56}{N}) are the uncertainties of the vehicle measurements and they compensate for the sensor noises. They are tuned by a statistical analysis of the vehicle sensor measurements, which consists of computing the standard deviation of the low-pass measured signal when the steering angle is null and the longitudinal velocity is constant \cite{bertipaglia2022two}. The variable $\sigma_{DD}$ is the level of distrust assigned to the pseudo-measurement $\beta_{DD}$ provided by CNN. The level of distrust computed by the CNN differs from a classic uncertainty measurement because it corresponds to the uncertainty of the pseudo-measurement scaled to match the weight of the noise parameters.

The observation model is used to perform the observability analysis. A sufficient condition to develop a UKF observer is the full rank condition of the observation model. A full rank equal to five is obtainable for all the operating regions in which $\delta \neq k\pi, \; \forall k\in \mathbf{Z}$ and $V_x\neq0$, where $\mathbf{Z}$ is the set of all the integers. The second condition is always respected because the measurement is considered when $V_x$ is higher than \SI{5}{m/s}. The steering angle is always inside the range $|\delta|\leq\pi/2$, so the only realistic unobservability happens when $\delta=0$. However, the vehicle sideslip angle is relevant for lateral dynamics, so it only happens when $\delta\neq0$.

\subsection{Training Phase}
\label{training}
\noindent The UKF-informed CNN is trained in a supervised way using a labelled dataset. The training is split into two phases: pre-training and end-to-end learning. 

\subsubsection{Pre-training} It consists of the back-propagation algorithm applied only to the CNN to speed up and stabilise the following end-to-end training phase. The sum of $\sigma_{V_y}$, $\sigma_{\dot\psi}$ MSE losses and $\beta_{DD}$, $\sigma_{DD}$ Gaussian negative log-likelihood loss constitute the cost function. The MSE loss functions ($MSE_{L,\,\sigma_{V_y}}$ and $MSE_{L,\,\sigma_{\dot\psi}}$) are represented as:
\begin{equation}
    \begin{split}
        & MSE_{L,\,\sigma_{V_y}} = \frac{1}{N}\sum_{i=1}^{N} \left( \hat\sigma_{V_y,\,i} - \sigma_{V_y,\,i} \right)^2\\
        & MSE_{L,\,\sigma_{\dot\psi}} = \frac{1}{N}\sum_{i=1}^{N} \left( \hat\sigma_{\dot\psi,\,i} - \sigma_{\dot\psi,\,i} \right)^2
    \end{split}
    \label{eq3}
\end{equation}
where $N$ is the size of the mini-batch (256), $\hat\sigma_{V_y}$ (\SI{0.0007}{m/s}) and $\hat\sigma_{\dot\psi}$ (\SI{0.002}{rad/s}) are the initial process model uncertainties tuned by the TSBO for the model-based approach. These losses steer the CNN to predict the process model uncertainties with a meaningful order of magnitude. The Gaussian negative log-likelihood loss function ($NLL_{L,\,\beta_{DD}}$) is represented in the following equation:
\begin{equation}
    \begin{split}
    &NLL_{L,\,\beta_{DD}}= \\
    &= \frac{1}{2}\sum_{i=1}^{N} \left( \log\left( \max\left( \sigma_{DD\,i}, \epsilon \right) \right)  + \frac{\left(\beta_{me,\,i}-\beta_{DD,\,i}\right)^2}{ \max\left( \sigma_{DD\,i}, \epsilon \right)}\right)
    \end{split}
    \label{eq4}
\end{equation}
where $\epsilon$ ($10^{-6}$) is a constant term for stability, and $\beta_{me}$ is the sideslip angle ground truth. The gradient used by the optimiser to update the CNN weights is not influenced by $\sigma_{DD}$. The pre-training leads the CNN's weights to estimate the correct $\beta$ with a meaningful $\sigma_{DD}$, they will be fine-tuned during the following end-to-end training phase. The sideslip angle ground truth is measured through the Corrsys-Datron optical speed sensor installed in the vehicle's front bumper. The sensor reference system is moved to correspond with the vehicle CoG. The measurement is filtered using a zero-phase low-pass filter (bandwidth \SI{5}{Hz}) because the training phase is sensitive to extreme outliers or noisy references \cite{viehweger2021vehicle}. The cost function is minimised by a mini-batch stochastic gradient descent algorithm based on a standard ADAM optimiser with a learning rate (0.0008). The training procedures' user-defined parameters are optimised through a Bayesian optimisation.

\subsubsection{End-to-end learning} It creates a mutualistic relationship between the model-based and data-driven approaches. The UKF is treated as a computation graph unrolled through time, so the CNN-UKF is discriminatively trained over the entire mini-batch length and not on a single step. The procedure to compute the loss function gradient is close to \cite{haarnoja2016backprop}, but in the proposed study a UKF is implemented rather than a linear Kalman filter. The first step is the computation of a loss a function ($L\left(\theta\right)$) that connects the output of the UKF-CNN ($\beta$) structure with the available ground truth ($\beta_{me}$). The training phase minimises the loss function error between the estimated and the measured sideslip angle, allowing to correctly estimate $\beta$ and all the parameters influencing it, so $\sigma_{DD}$, $\sigma_{\dot{\psi}}$ and $\sigma_{V_y}$.
The loss function depends on the CNN's weights ($\theta$) and it is based on the following equation:
\begin{equation}
    L\left(\theta\right) = \frac{1}{N}\sum_{i=1}^{N}\left( \beta_{me,\,i} - \beta_{DD,\,i} \right)^2 + \frac{1}{N}\sum_{i=1}^{N}\left( \beta_{me,\,i} - \beta_i \right)^2
    \label{eq5}
\end{equation}
where $\beta_{DD}$ is the output of the CNN and $\beta_{me}$. The first loss function part, $\frac{1}{N}\sum_{i=1}^{N}\left( \beta_{me,\,i} - \beta_{DD,\,i} \right)^2$, involves just one CNN output and it is not affected by the UKF. It helps the CNN to estimate the correct pseudo-measurement $\beta_{DD}$. The second part of the loss function, $\frac{1}{N}\sum_{i=1}^{N}\left( \beta_{me,\,i} - \beta_i \right)^2$, is affected by all four CNN outputs and by the UKF. The second step to train the proposed UKF-CNN is the computation of the gradient of the loss function with respect the CNN weights ($\nabla_\theta L\left(\theta\right)$). This is performed following the typical BPTT algorithm. Moving backward from the loss function, the $\nabla_\theta L\left(\theta\right)$ is computed by a recursive computation of the loss function gradient with respect to the vehicle states from $t-1$ to $t$ according to:
\begin{equation}
    \frac{\partial L}{\partial x_{s,\,t-1}} = \frac{\partial {ukf}_{t-1}}{\partial x_{s,\,t-1}}\frac{\partial L}{\partial {ukf}_{t-1}} + \frac{\partial x_{s,\,t}}{\partial x_{s,\,t-1}}\frac{\partial L}{\partial x_{s,\,t}}
    \label{eq6}
\end{equation}
where ${ukf}_{t-1}$ represents all the functions that describes the UKF algorithm, i.e. process model, observation model and Kalman gain computation. The output of the UKF algorithm depends from the vehicle states, inputs, measurements and from the CNN outputs. The gradient computation continues applying the chain rule to eq. \ref{eq6}, and moving backward computing the derivative with respect to each CNN weight, as for a normal NN. The training is based on a mini-batch stochastic gradient descent algorithm (mini-batch size of 256) based on a standard ADAM optimiser with a learning rate (0.0008) is implemented.

\section{Experiment Setup}
\noindent This section describes how the experiments have been conducted and how the proposed approach has been compared to the baseline methods.
\label{exp}
\subsection{Dataset}

\noindent The experiments have been conducted at the Automotive Testing Papenburg GmbH with the test platform based on a BMW Series 545i. The vehicle was instrumented with the standard IMU, wheel force transducers and intelligent bearings for each wheel, GPS and a Corrsys-Datron optical sensor to measure the sideslip angle (measurement accuracy of \SI{\pm0.2}{\degree}). The high-end optical speed sensor is used to measure the ground truth. The intelligent bearings demonstrate a similar accuracy to the wheel force transducer \cite{kerst2016reconstruction}, the most common sensor technique in research for tyre force measurement. Thus, the tyre forces in the training dataset are taken from the wheel force transducers making the paper easier to reproduce.
The dataset contains 216 manoeuvres corresponding to two hours of driving and consists of standard vehicle dynamics manoeuvres, e.g. double lane change, slalom, random steer, J-turn, spiral, braking in the turn, and steady-state circular tests, together with recorded laps at the handling track. All manoeuvres were driven on dry asphalt with tyres inflated according to the manufacturer's specifications. The bank angle and the road slope were negligible, and the friction coefficient was approximately constant. Two different electronic stability control settings (On, Off) were used. All the measurements were recorded at \SI{100}{Hz}, the standard frequency for vehicle state estimation. A statistical outlier removal has been applied to remove extreme outliers. However, particular attention is paid not deleting edge case measurements which are the most important data. Furthermore, all the manoeuvres were manually inspected to check the outlier removal efficacy. The measurements are considered when $V_x$ is higher than \SI{5}{m/s} and are filtered using a low-pass zero-phase filter with a cut-off frequency of \SI{5}{Hz} based on a finite impulse response technique \cite{viehweger2021vehicle}.

\begin{figure}[!t]
    \centering
    \includegraphics[height=3cm, keepaspectratio]{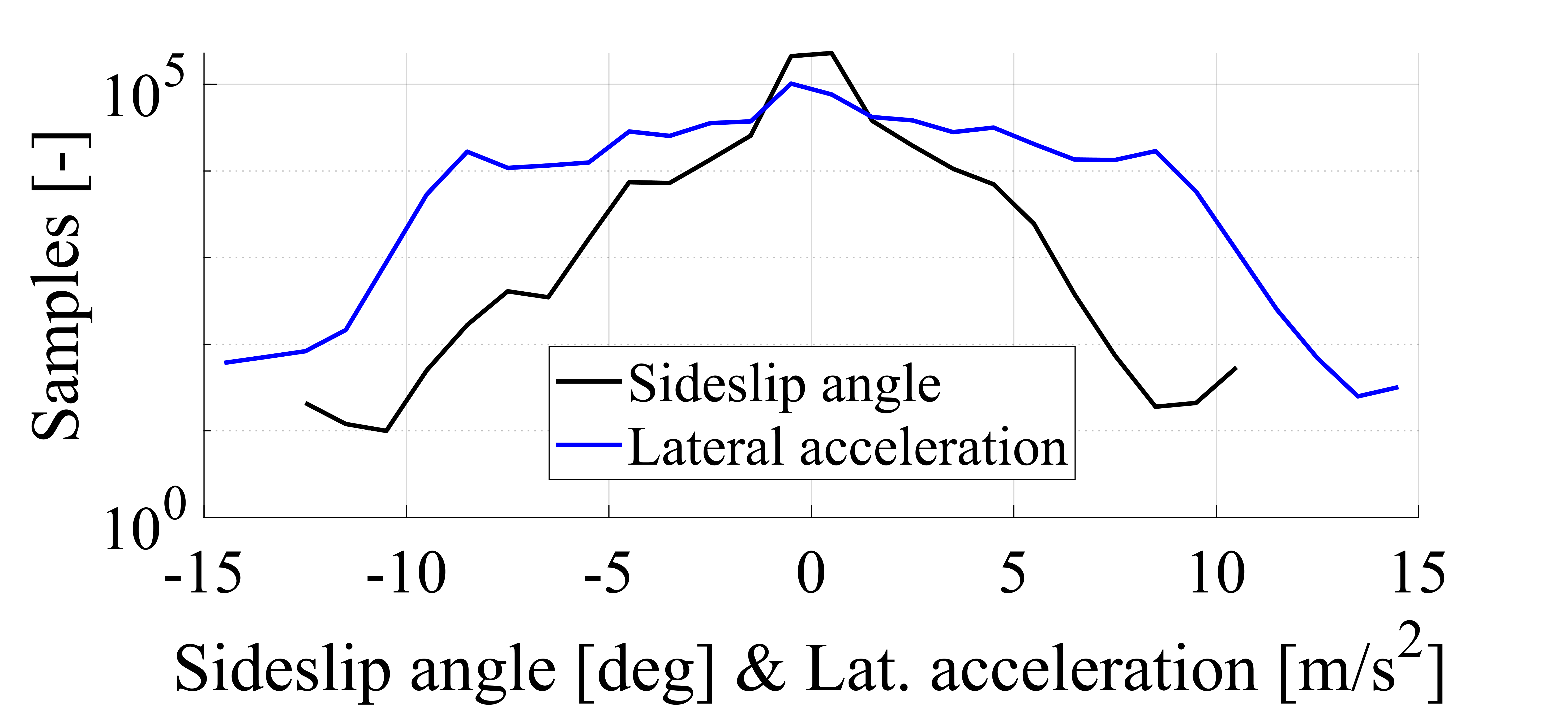}
    \caption{Log distribution of sideslip angle and lateral acceleration. Each bin corresponds to \SI{1}{deg} and \SI{1}{m/s^2}}
    \label{fig:logdensity}
\end{figure}

The log distribution of the sideslip angle and lateral acceleration is represented in Fig. \ref{fig:logdensity}. The lateral acceleration is almost spread equally in the range $[-10,\,10]$ $\mathrm{m/s^2}$. In contrast, the sideslip angle measurements mainly distribute in the range $[-3,\,3]$ deg. The latter is a common phenomenon because it is challenging to perform manoeuvres with a high sideslip angle, even when the vehicle has a very high lateral acceleration. Especially in dry road conditions, only a professional driver can induce a high sideslip angle.

A second dataset is selected from the same measurements. It will be referenced as limited dataset because it only contains measurements of when the vehicle has lateral acceleration $|a_y|\leq 7\,\mathrm{m/s^2}$. This simulates the cost and complexity of recording a large number of manoeuvres in which the vehicle is driven at the extreme vehicle behaviour, but not at the handling limits. Such a situation is common in the automotive field because, at the handling limits, the driver can easily lose the vehicle's control. Thus, the limited dataset will be used to analyse the proposed hybrid approach regarding its robustness and generalisation capabilities.

Both datasets are split into three sub-sets: training (\SI{75}{\%}), validation (\SI{15}{\%}) and test (\SI{10}{\%}). The test set contains the same manoeuvres for both the full and limited datasets. It consists of manoeuvres representing the entire driving behaviour, but more focus is paid to highly non-linear situations. It includes 23 manoeuvres: two braking in the turn, two skidpad, five J-turn, four slalom, four lane change, two random steers, three spiral and one lap track.

\subsection{Key Performance Indicators}

\noindent The performance of the different approaches is assessed through four key performance indicators (KPIs), which are commonly used in sideslip angle estimation \cite{mazzilli2021benefit,bertipaglia2022model,doumiati2010onboard}.
\begin{itemize}
    \item The MSE assesses the overall estimation performance.
    \item The non-linear MSE (MSE\textsubscript{nl}) corresponds to the MSE computed only when $|a_y|\geq 4\,\mathrm{m/s^2}$. It measures the estimation performance when the vehicle behaves non-linearly.
    \item The absolute maximum error (ME) measures the worst estimation performance.
    \item The non-linear ME (ME\textsubscript{nl}) measures the worst estimation performance in the case of non-linear vehicle behaviour.
\end{itemize}
The non-linear KPIs analyse the hybrid approach performance in the most critical scenarios. The MSE and MSE\textsubscript{nl} are used to evaluate the estimation accuracy, while  ME and ME\textsubscript{nl} are used to assess temporary high errors in the estimation. The latter is relevant to assess whether the estimation is always coherent with the physical vehicle behaviour.

\subsection{Baseline Methods}

\noindent The proposed hybrid approach is compared with the state-of-art model-based, data-driven and hybrid approaches. All the considered baselines are adapted and optimised to use the same sensor setup and dataset, ensuring an objective and fair comparison.

The model-based approach is a UKF-based on a single-track model with tyre force measurements, as presented in \cite{bertipaglia2022model}. The process noise parameters are tuned with the TSBO, and the observation noise parameters associated with the tyre force measurements are adapted online to enhance the observer's performance. The adaptability is related to the reduction of the level of noise coupled with tyre force measurements, this increases the Kalman gain when the vehicle behaves non-linearly. Thus, the effect of the Kalman gain is magnified during manoeuvres at the handling limit. Otherwise, a magnified Kalman gain when the vehicle behaves linearly could influence the vehicle states to follow the measurement sensor noises. The adaptability is triggered with a hysteresis loop to avoid the chattering phenomenon.

The data-driven approach is a FFNN that uses IMU and tyre force measurements as inputs, as evaluated in \cite{bertipaglia2022model}. A simple FFNN reaches a better performance than a RNN when the tyre force measurements are included in the input set because the RNN prediction power is insufficient to compensate for the higher numbers of parameters to be trained. The NN is formed by two hidden layers with respectively 250 and 125 neurons each and ReLU activation functions. It uses a dropout regularisation technique (0.2) and a Xavier initialisation to avoid overfitting. An early stopping method with patience equal to 20 is applied for the same reason. The MSE is the loss function minimised by a mini-batch stochastic gradient descent algorithm based on a standard ADAM optimiser with a learning rate (0.001). The mini-batch size is 1024. For the training procedures, user-defined parameters are optimised through a Bayesian optimisation.\compressParag

The hybrid approach is a deep ensemble-UKF (DE-UKF) \cite{kim2020vehicle} adapted to maximise the estimation performance on a dataset with tyre force measurements. The DE is formed by 20 FFNNs trained independently on the same dataset. The FFNNs different estimations are combined in a model averaging. Hence, the final $\beta_{DD}$ is the mean of the FFNNs estimations, and $\sigma_{DD}$ is the variance of the different model estimations. Each FFNN is trained using a Gaussian negative log-likelihood cost function optimised through mini-batch stochastic gradient descent based on an ADAM optimiser with a learning rate (0.0008). The epoch's number for each FFNN is 30. DE relies on the stochasticity of neural network training, which allows every FFNN to converge to a different set of parameters. However, the estimation accuracy is low when all models predict incorrectly, and there is no guarantee that the $\sigma_{DD}$ will be high. This especially happens when the error is in the low sideslip angle range because the NNs estimations tend to be closer. A high level of distrust suggests that the UKF does not rely on the data-driven pseudo-measurement but trusts the estimation of the UKF process model. Vice-versa, when the level of distrust is low, the UKF considers the neural network estimation reliable. $\sigma_{DD}$ must be scaled before being used by the UKF because the output of the DE does not match the weight of the other noise parameters. Otherwise, the UKF puts too much trust in $\beta_{DD}$. The scaling is based on an exponential function (eq. \ref{eq:scaling}) which differentiates approximately similar $\sigma_{DD}$.\compressParag
\begin{equation}
    \sigma_{DD, \,sc} = 10^{p_1} \sigma_{DD}^{p_2}
    \label{eq:scaling}
\end{equation}
where $p_1$ (-4.2690 for the full dataset and -1.4353 for the limited one) and $p_2$ (0.7901 for the full dataset and 1.465 for the limited one) are two scaling parameters tuned using a Bayesian Optimisation. The values of $p_1$ and $p_2$ change according to the dataset because they strongly influence the DE's estimation performance. If $p_1$ and $p_2$ are not re-tuned for the limited dataset, the UKF will put too much trust in the DE, even if it lacks performance.

\section{Results}
\label{res}
\noindent This section demonstrates the performance of the proposed approach. Subsection \ref{subsres} analyses how accurate the proposed approach is with respect to the baselines when it is trained using a full dataset. Subsection \ref{subsres2} shows the results of the robustness analysis when only a limited dataset is available. This demonstrates that the data-driven approach is highly influenced by the amount and quality of the data.

\subsection{Full Dataset Results}
\label{subsres}
\noindent The CNN-UKF, the DE-UKF and the data-driven approach have been trained using the full dataset.
\begin{table}[htp]
    \caption{Sideslip angle estimation comparison using the full dataset.}
    \label{tab:Optim_Data_Comp}
    \begin{center}
    \begin{tabular}{ >{\centering\arraybackslash}m{1in}  >{\centering\arraybackslash}m{0.3in}  >{\centering\arraybackslash}m{0.3in} >{\centering\arraybackslash}m{0.3in}  >{\centering\arraybackslash}m{0.3in} }
    \toprule[1pt]    
    \textbf{Approaches} &  \textbf{MSE [\SI{}{deg^2}]}  &  \textbf{MSE\textsubscript{nl} [\SI{}{deg^2}]} &  \textbf{ME [\SI{}{deg}]} &  \textbf{ME\textsubscript{nl} [\SI{}{deg}]}\\
    \hline    
    Model-based         & 0.161             & 0.277             & 1.111             & 0.991 \\
    Data-driven         & 0.096             & 0.157             & 1.293             & 1.123 \\
    DE-UKF              & 0.087             & 0.157             & 0.981             & 0.822 \\
    \textbf{CNN-UKF}    & \textbf{0.086}    & \textbf{0.118}    & \textbf{0.979}    & \textbf{0.776} \\
    \bottomrule[1pt]
    \end{tabular}
    \end{center}
\end{table}

The overall comparison is presented in Table \ref{tab:Optim_Data_Comp}. Both hybrid approaches perform better than the model-based and data-driven approaches considering all four KPIs. This highlights the importance of the hybrid architecture for vehicle sideslip angle estimation. For instance, the model-based approach has a higher MSE and MSE\textsubscript{nl} than the data-driven approach but a lower average ME and ME\textsubscript{nl}. The hybrid approaches have the same estimation accuracy (MSE and MSE\textsubscript{nl}) as the data-driven approach without the average higher ME.  The reason is that in a hybrid approach, data-driven estimation is always validated through the model-based approach. 

It can be seen that CNN-UKF outperforms the three other approaches for all the proposed KPIs. However, it does not have the same benefits in magnitude for all of them. The overall MSE and ME of DE-UKF and CNN-UKF are comparable. The minor improvements for the linear vehicle behaviour are respectively \SI{1.15}{\%} and \SI{0.20}{\%} in favour of the CNN-UKF. Anyhow, if the performance is evaluated when the vehicle behaves non-linearly, the CNN-UKF will strongly outperform DE-UKF with an improvement of \SI{24.84}{\%} for the MSE\textsubscript{nl} and \SI{5.60}{\%} for the ME\textsubscript{nl}. A possible explanation is that the end-to-end training informs CNN about the vehicle dynamics compensating for the lower amount of data in this operating condition. 

On the contrary, the DE during the training is not aware of the physical vehicle behaviour, so it is subjected to a decay in performance where the dataset has fewer samples. The DE becomes aware of the UKF performance only while tuning the level of distrust scaling parameters. Furthermore, the process model noise parameters are online adapted in the CNN-UKF, allowing the UKF to accommodate better the mismatches between the physical and modelled vehicle behaviour.

\begin{figure}[!t]
    \centering
    \includegraphics[height=3.5cm, keepaspectratio]{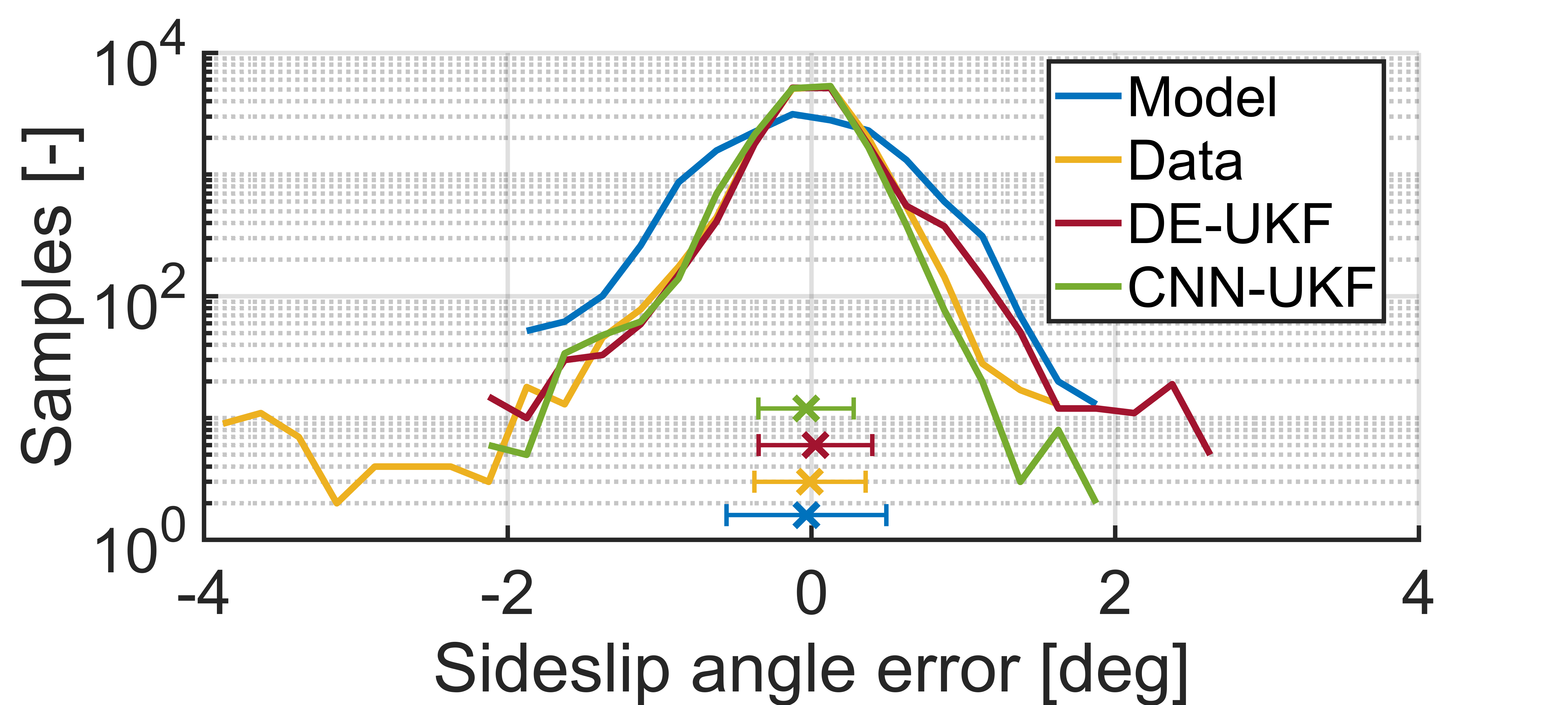}
    \caption{Distribution of the sideslip angle error when the vehicle $|a_y|>$ \SI{4}{m/s^2} for every approach in the test set. Each bin is \SI{0.25}{deg} wide. The x represents the mean and the line between the vertical symbols ($|-|$) is the standard deviation of the sideslip angle error.}
    \label{fig:Dist_Opt_Non}
\end{figure}

Similar conclusions can be stated from the log distribution of the sideslip angle error in the non-linear operating range, see Fig. \ref{fig:Dist_Opt_Non}. The data-driven and the hybrid approaches have a similar amount of $\beta$ error samples in the range $\left[-1.5,\,1.5\right]$ deg. In contrast, the model-based approach suffers from the lower accuracy of the vehicle model in the non-linear operating region. However, the data-driven approach and partially the DE-UKF are more prone to high estimation errors (\SI{\geq1.5}{deg}) than the model-based and CNN-UKF. The latter outperforms all other approaches and has the $\beta$ error mean closest to zero and the lowest standard deviation. Hence, a UKF coupled with a data-driven approach has the same performance as a data-driven approach in a low error range, but it reduces the sporadic high errors of a purely data-driven approach. Furthermore, the end-to-end training and the process noise parameters adaption allow the CNN-UKF to maximise the hybrid capability especially when the vehicle $|a_y|>$ \SI{4}{m/s^2}.

\begin{figure}[!t]
    \centering
    \includegraphics[height=3.5cm, keepaspectratio]{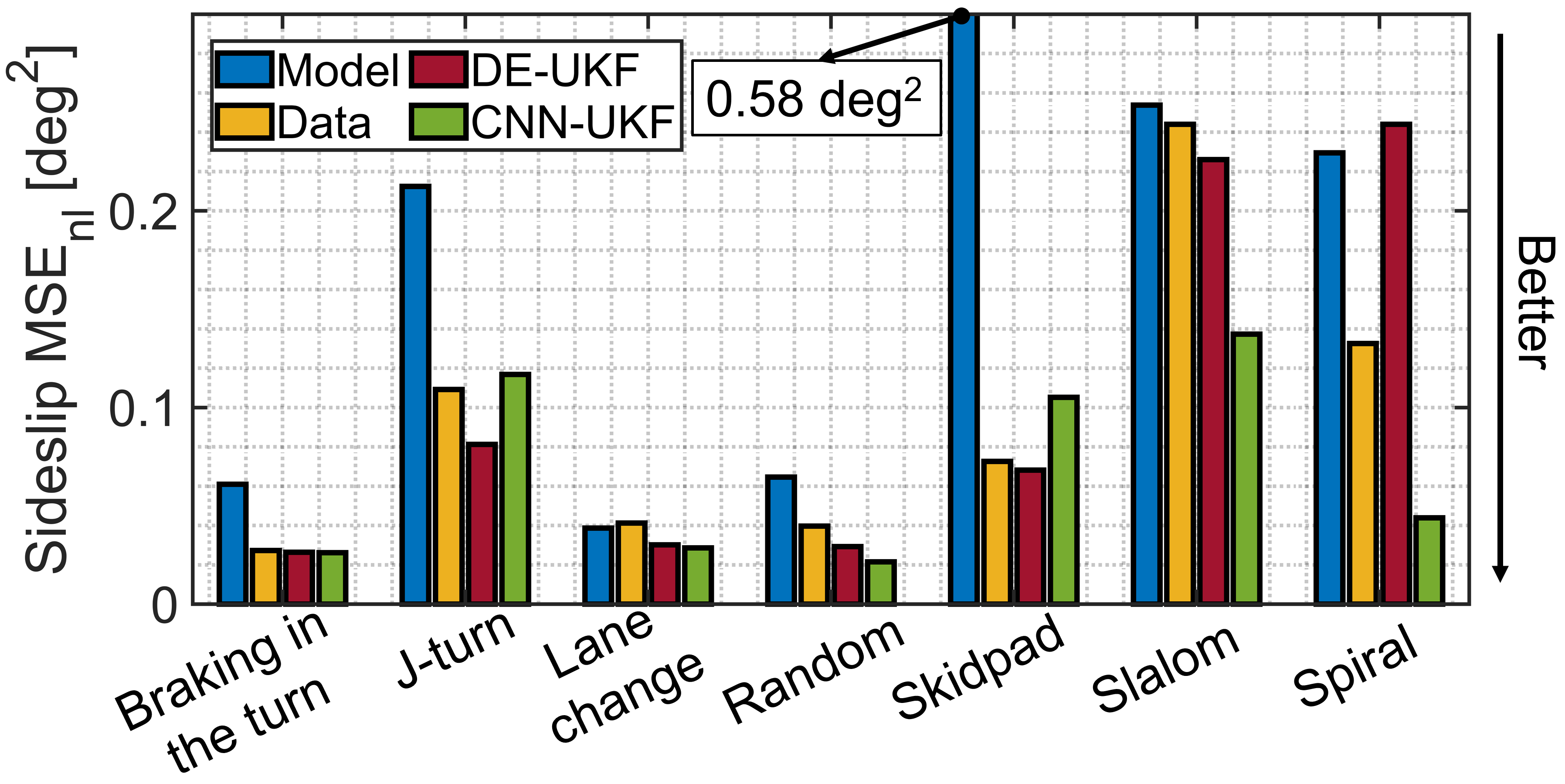}
    \caption{Sideslip angle MSE\textsubscript{nl} comparison for every group of manoeuvres.}
    \label{fig:Opt_bar}
\end{figure}

\begin{figure}[!t]
    \centering
    \includegraphics[height=3.5cm, keepaspectratio]{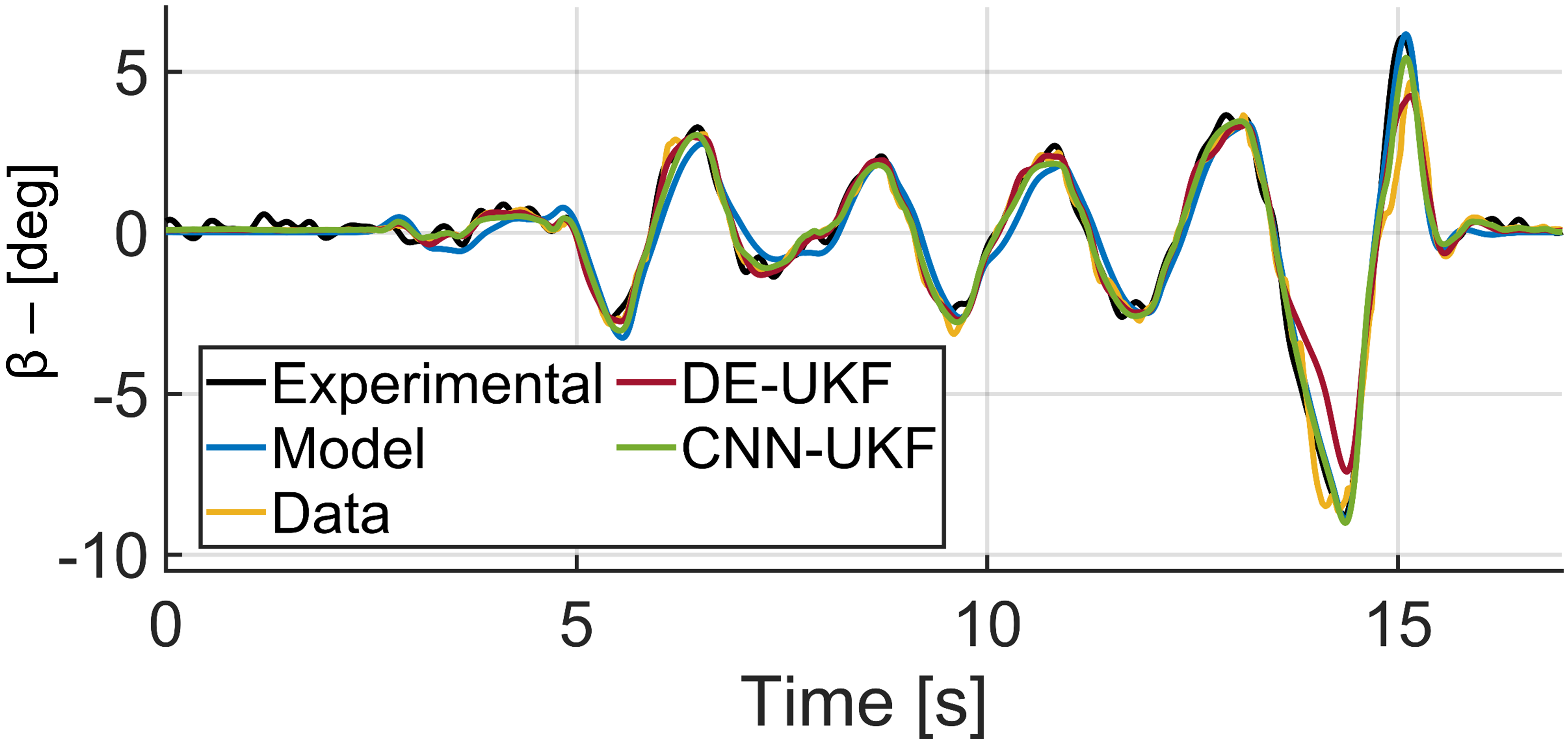}
    \caption{Slalom manoeuvre. Comparison of the sideslip angle estimation between all four approaches.}
    \label{fig:slalom}
\end{figure}

Fig. \ref{fig:Opt_bar} analyses how the estimation performance change for different manoeuvres. The model-based approach has a weak accuracy, especially in braking-in-the-turn, J-turn and skidpad tests. The braking-in-the-turn involves a coupling between the longitudinal and lateral dynamics, which is not modelled in the used single-track vehicle model. In a J-turn manoeuvre, the vehicle is driven at the limits of handling, where the mismatches between the physical and modelled vehicle behaviour are higher. Whereas for skidpad tests, the explanation is that it is a quasi steady-state manoeuvre, so the vehicle yaw acceleration is almost null, and the difference between estimated and measured tyre forces becomes essential for the $\beta$ estimation. The tyre model is one of the most significant uncertainty sources in the model-based approach. The data-driven approach almost constantly behaves better than the model-based but worse than the hybrid approaches  for estimation accuracy. However, it outperforms the DE-UKF in a spiral manoeuvre, and a possible explanation is that the DE-UKF puts too much trust in the UKF process model. The CNN-UKF outperforms all the other approaches in five out of seven manoeuvres. Particularly relevant is the improvement in the slalom and spiral manoeuvres. The slalom has the highest number of sideslip angle peaks (Fig. \ref{fig:slalom}), which are the most difficult moments to estimate sideslip. Spiral manoeuvres are particularly challenging because it has an extra turn respect the J-turn.

\begin{figure*}[!t]
    \centering
    \hspace*{0.5cm}
    \subfloat[\label{fig:Slalom_A}]{%
       \includegraphics[height=3.5cm, keepaspectratio]{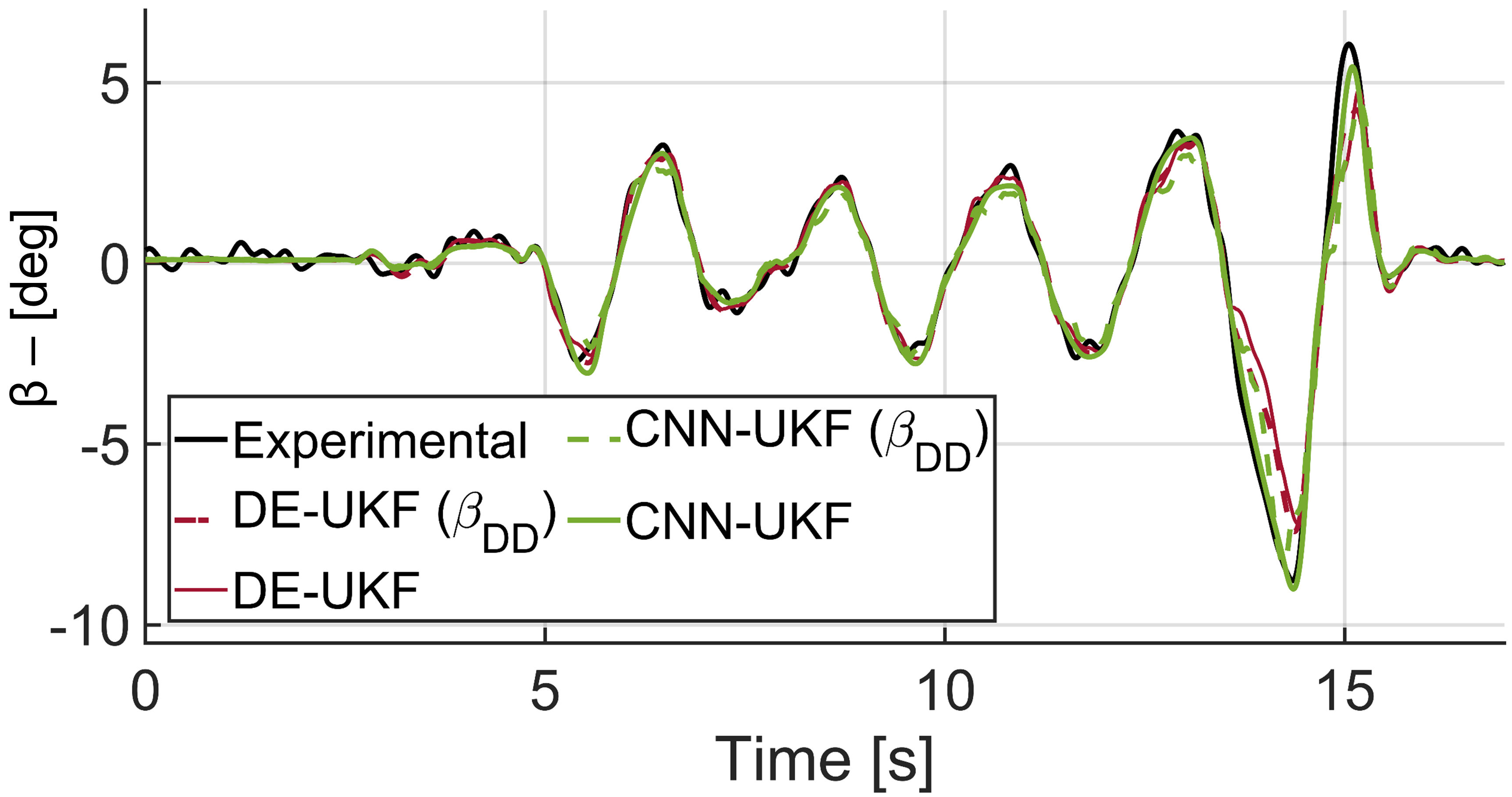}} \hfill
    \subfloat[\label{fig:Slalom_B}]{%
       \includegraphics[height=3.5cm, keepaspectratio]{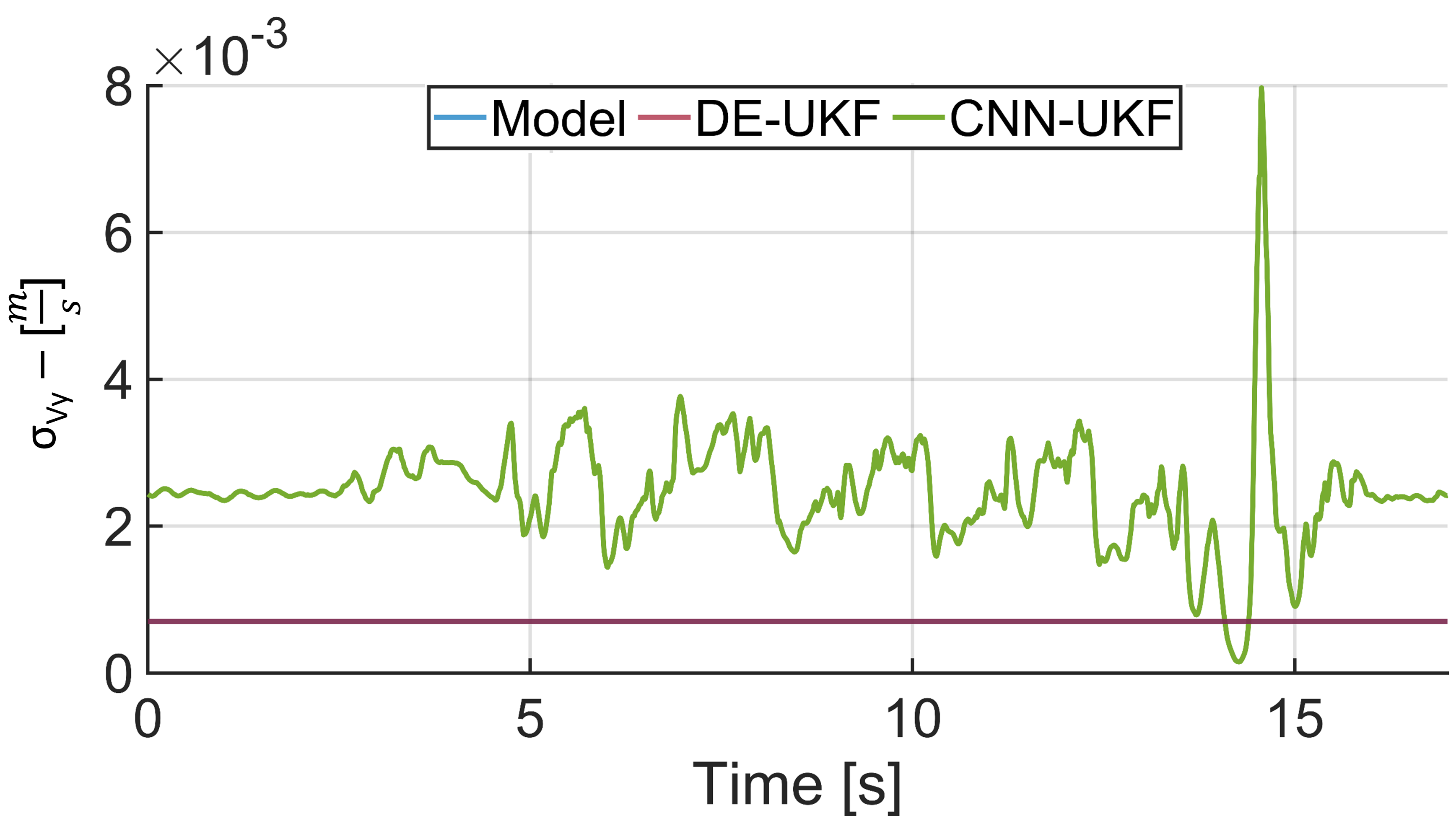}}\hspace*{0.5cm}
       
    \hspace*{0.5cm}
    \subfloat[\label{fig:Slalom_C}]{%
       \includegraphics[height=3.5cm, keepaspectratio]{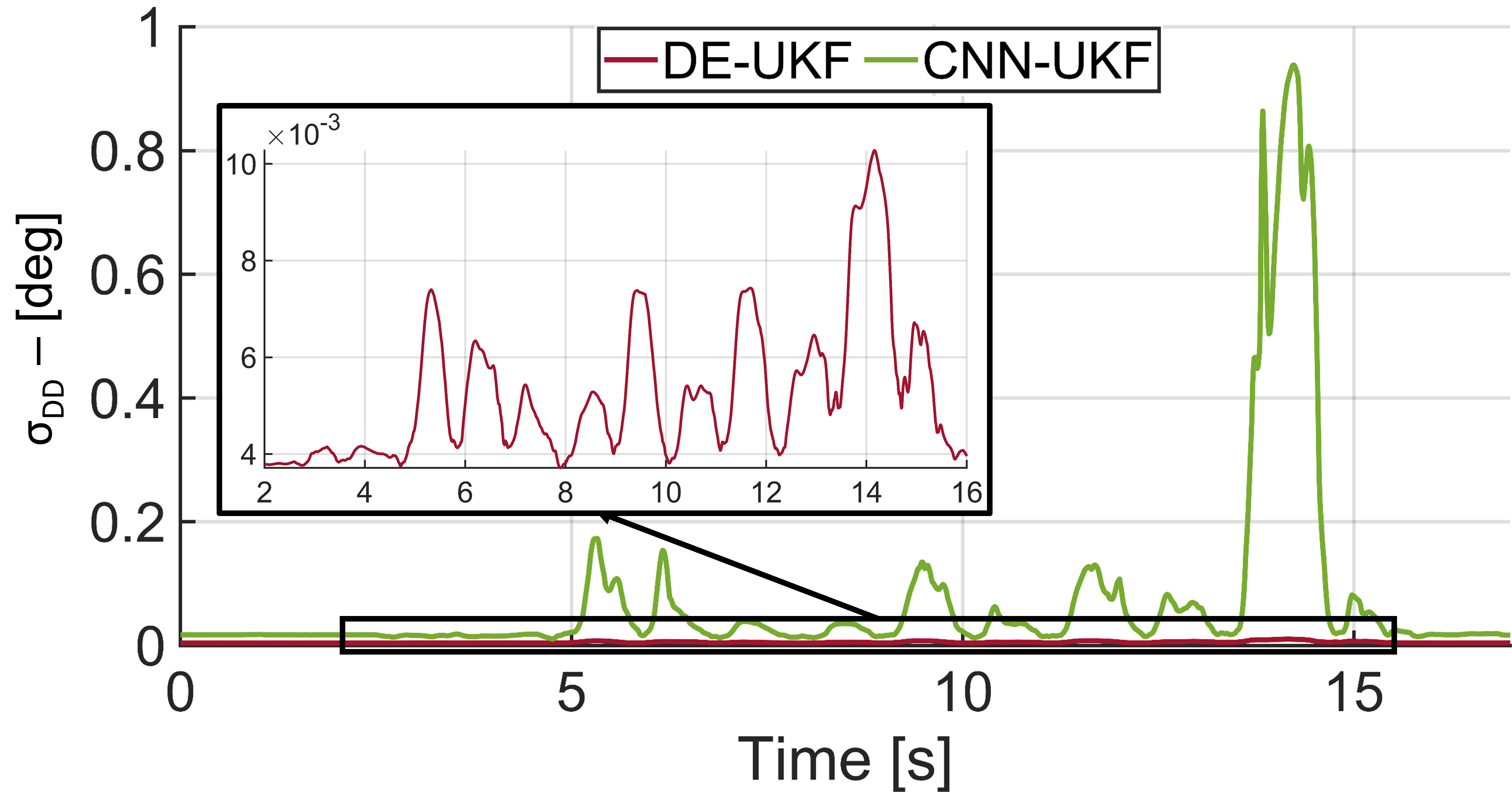}} \hfill
    \subfloat[\label{fig:Slalom_D}]{%
       \includegraphics[height=3.5cm, keepaspectratio]{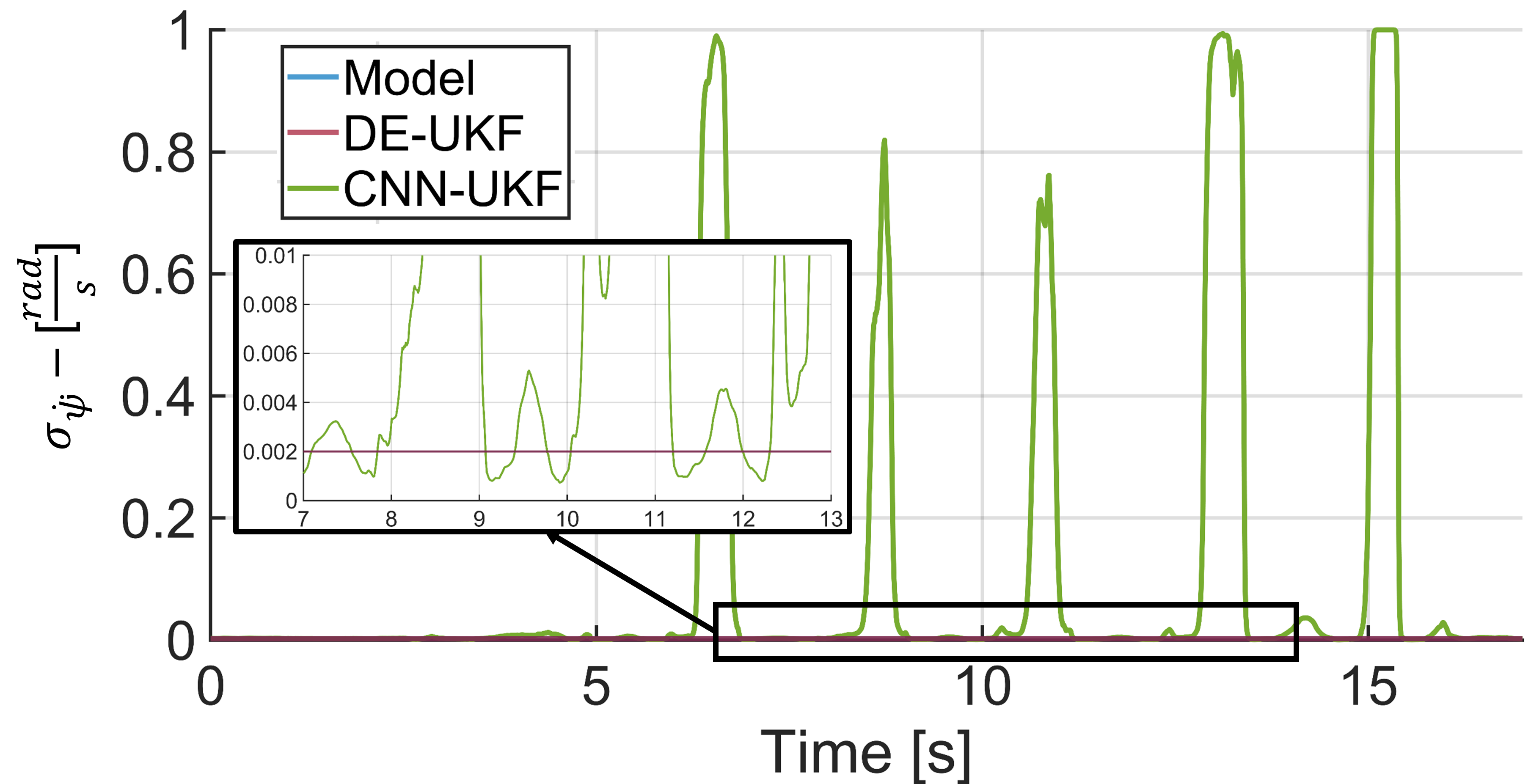}} \hspace*{0.5cm}
    \caption{Fig: \ref{fig:Slalom_A} shows the estimated and the pseudo-measurement of the sideslip angle. Fig: \ref{fig:Slalom_B} shows the process noise parameter associated with the $V_y$. Fig. \ref{fig:Slalom_C} shows the level of distrust in in the NN for the hybrid approaches. Fig. \ref{fig:Slalom_D} shows the process noise parameter associated with the $\dot{\psi}$.}
    \label{fig:slalom_data}
\end{figure*}

Fig. \ref{fig:slalom} shows the sideslip angle estimation in a slalom manoeuvre at the handling limits. All four approaches provide a reliable estimation, but the CNN-UKF outperforms the other approaches when the vehicle reaches a $\beta$ peak of \SI{10}{deg} at around \SI{14}{s}. This is a typical situation where a correct estimation of $\beta$ is essential to help the vehicle control system maintain vehicle stability. Thus, an improved estimation in this condition is particularly relevant for safety. The already mentioned high non-linearities reduce the accuracy of the model-based approach. The data-driven approach lacks accuracy at \SI{15}{s} due to the few data in the training set describing this vehicle's operation point. The DE-UKF improves the estimation performance between \SI{5}{s} and \SI{13}{s} combining the pros of the model-based and data-driven approach, but it lacks performance at around \SI{14}{s}. CNN-UKF improves the estimation accuracy not only in the range of $[5,\;10]$ s but also in the highest peak at \SI{14}{s}, as can be observed in Fig. \ref{fig:slalom_data}. \compressParag

Fig. \ref{fig:Slalom_A} shows the $\beta$ and $\beta_{DD}$ for the hybrid approaches. CNN-UKF and DE-UKF $\beta_{DD}$s lack accuracy between \SI{12}{s} and \SI{15}{s}, but the CNN-UKF $\beta$ is accurate because the UKF is correctly weighting the UKF process model's information with the NN's pseudo-measurement. Vice-versa, the UKF of the DE-UKF puts too much trust in $\beta_{DD}$. When the $\beta_{DD}$ error rises, the corresponding level of distrust (Fig. \ref{fig:Slalom_C}) also grows. CNN-UKF and DE-UKF $\sigma_{DD}$s have the same order of magnitude in normal driving, but the one related to CNN-UKF rises much more than the DE-UKF. This broader range makes the proposed approach much less confident in the NN when its output is incorrect. This is not possible for the DE-UKF due to its training process. The DE-UKF does not have end-to-end training, so its $\sigma_{DD}$ cannot match the weight of the other UKF noise parameters. The DE-UKF $\sigma_{DD}$ non-linear scaling compensates only partially this issue. Fig. \ref{fig:Slalom_C} clearly demonstrates how the CNN-UKF distrust level range is $\left[10^{-3},\, 1\right]$, while the range for the DE-UKF is only $\left[10^{-3},\, 10^{-2}\right]$.\compressParag

Another explanation for the better performance of the CNN-UKF is related to the online adaptation of the process noise parameters. The adaptive parameters allow the UKF to know the current mismatches between the modelled and physical vehicle behaviour. The process noise parameters of the DE-UKF and model-based approach are constant, so they correspond to a trade-off between the different driving conditions. Vice-versa, the CNN-UKF relies on optimal tuned process noise parameters every instant. Fig. \ref{fig:Slalom_B} and \ref{fig:Slalom_D} show the values of $\sigma_{V_y}$ and $\sigma_{\dot\psi}$, respectively. As expected from the literature \cite{acosta2017optimized}, both increase with the growth of vehicle non-linearities. This further proves that CNN-UKF behaves according to physical vehicle motion. $\sigma_{V_y}$ has a peak at \SI{14}{s}, corresponding to the last vehicle's right turn, where the rear inner tyre is even detached from the ground due to the aggressiveness of the manoeuvre. This extreme condition is created by a transient lateral load transfer (not modelled) which strongly influences lateral tyre force production, resulting in a significant $V_y$ model mismatch. Moreover, the effect of the front axle longitudinal force ($F_{xf}$) on the lateral velocity is not modelled $\left(\frac{F_{xf}\sin\left(\delta\right)}{m}\right)$. Overall, the constant and the adapted process noise parameter have the same magnitude. Still, the one associated with CNN-UKF is generally bigger (apart from \SI{13}{s} to \SI{14}{s}). The reason is that the constant $\sigma_{V_y}$ was optimised, considering also less aggressive manoeuvres where the vehicle model is more reliable.

The process noise parameter $\sigma_{\dot\psi}$ rises by two orders of magnitude when the vehicle has a high sideslip angle. At the same time, when $\beta$ is low, the adapted $\sigma_{\dot\psi}$ is slightly lower than the constant process noise parameter. A possible explanation is that the mismatches of the modelled $\dot\psi$ are higher than that of $V_y$. The meaningful adaptability of the process parameter noises shows the CNN-UKF has an insight into vehicle dynamics physics and it can online compensate for it.\compressParag
\begin{figure}[!t]
    \centering
    \includegraphics[height=3.5cm, keepaspectratio]{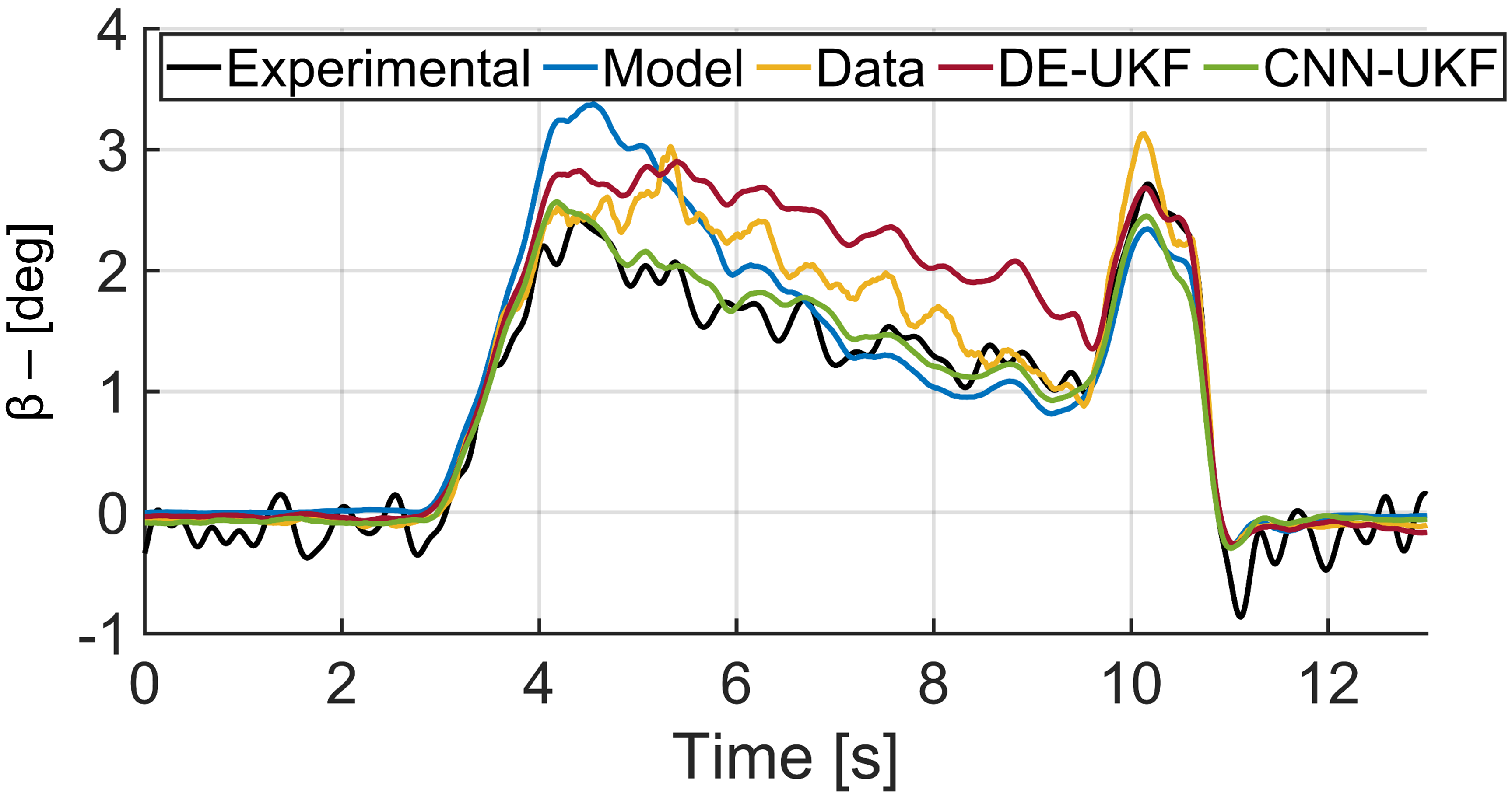}
    \caption{Spiral manoeuvre. Comparison of the sideslip angle estimation between all four approaches.}
    \label{fig:spiral}
\end{figure}
\begin{figure}[!t]
    \centering
    \subfloat[\label{fig:Hock_Sideslip}]{%
       \includegraphics[height=3.7cm, keepaspectratio]{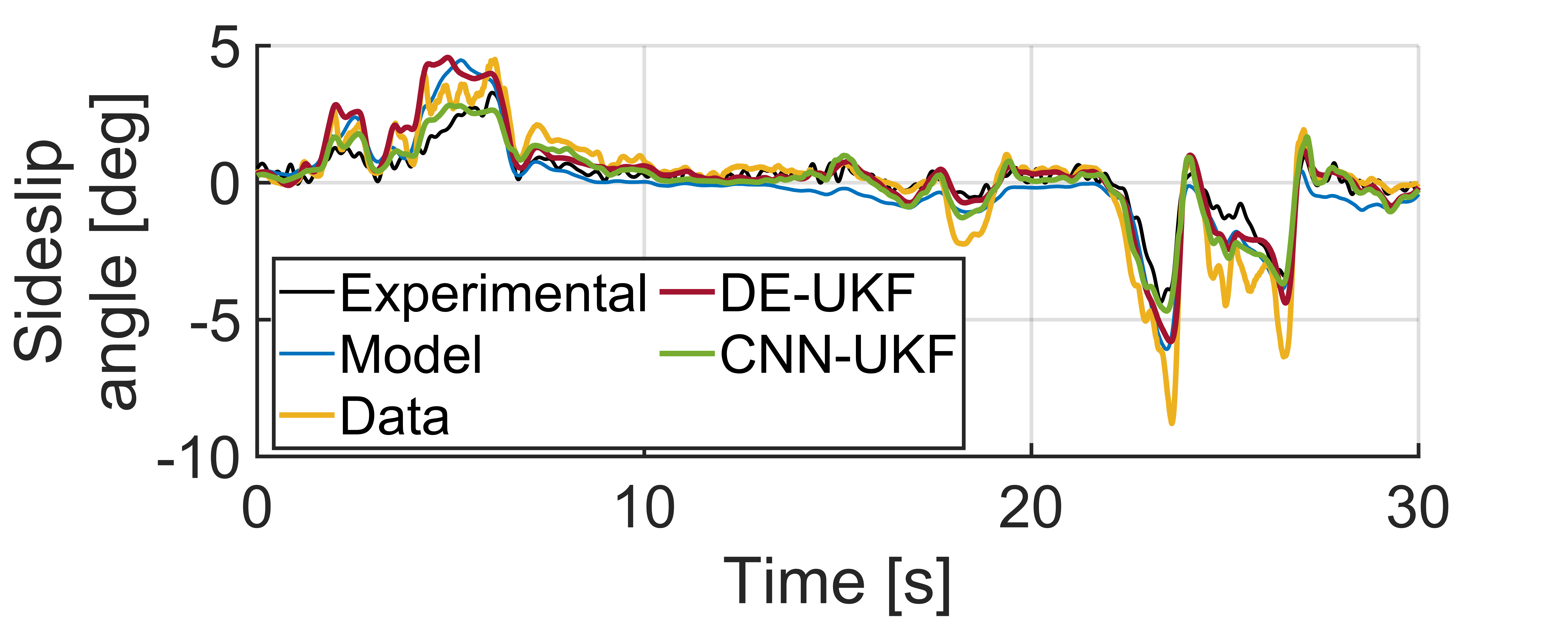}} \hfill 
        \subfloat[\label{fig:Hock_Acc}]{%
       \includegraphics[height=3.5cm, keepaspectratio]{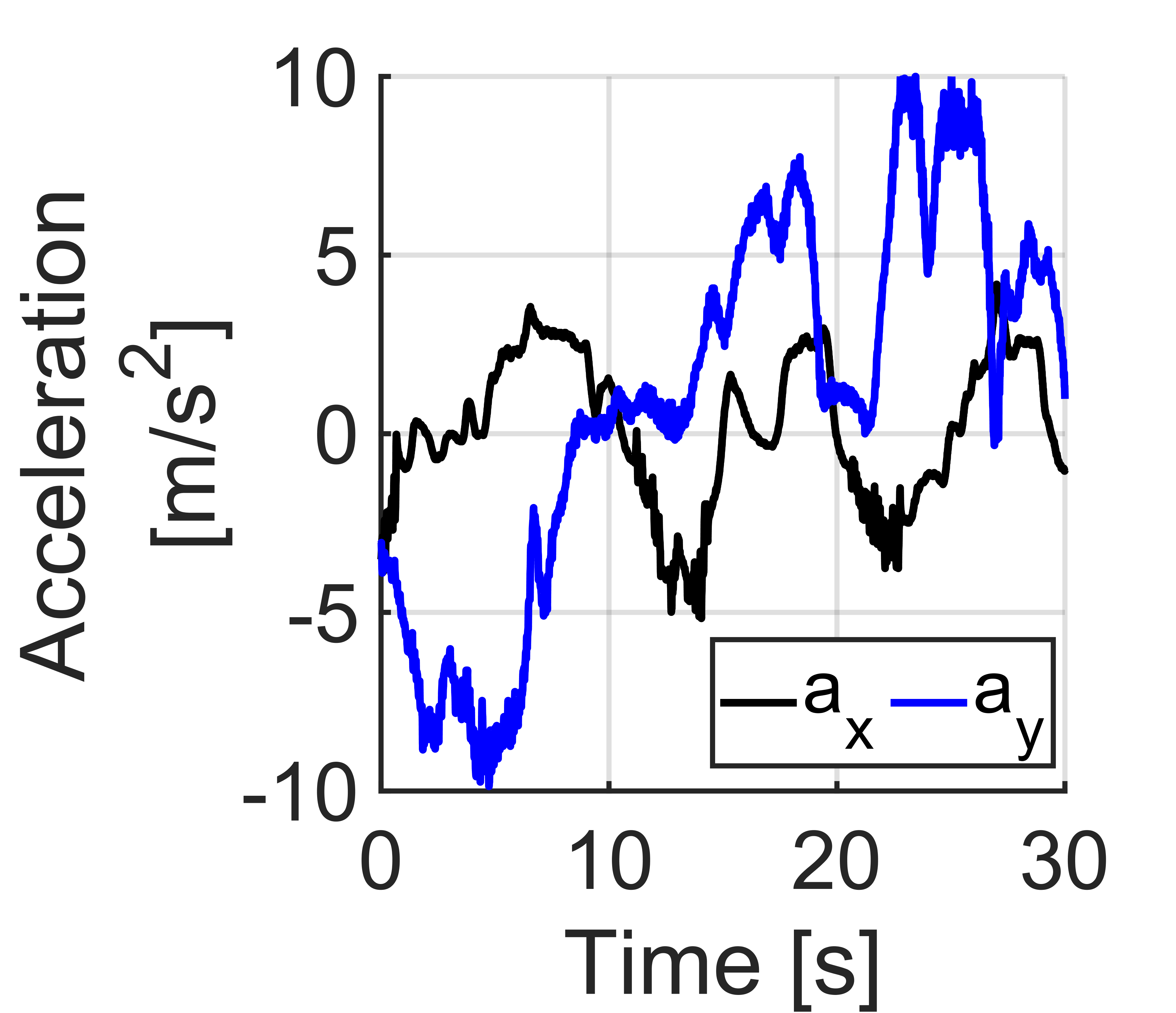}} 
    \subfloat[\label{fig:Hock_DD}]{%
       \includegraphics[height=3.5cm, keepaspectratio]{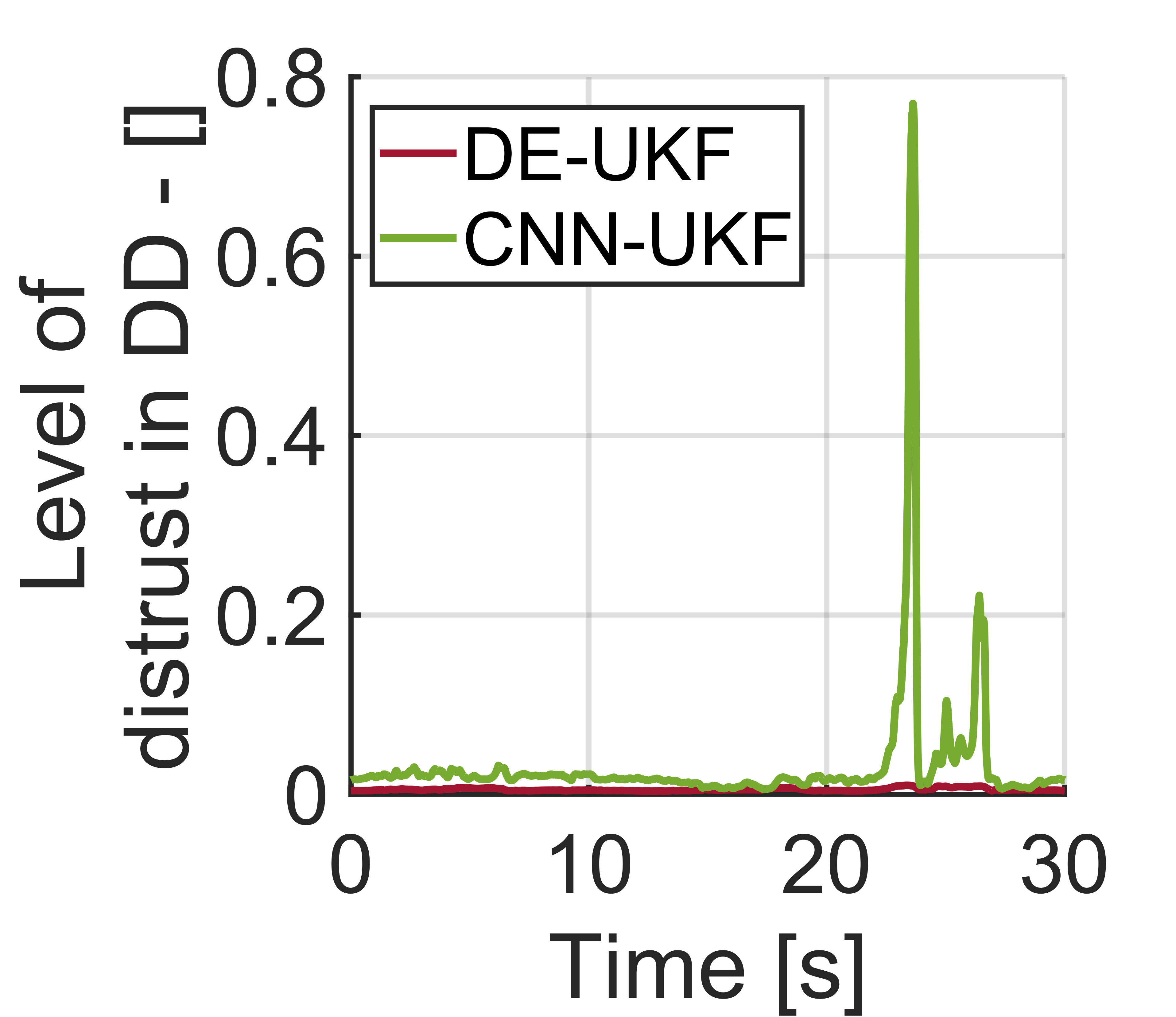}} \hfill
    \caption{Fig. \ref{fig:Hock_Sideslip} compares the sideslip angle estimation between four approaches in a portion of a racing track. Fig. \ref{fig:Hock_Acc} shows the recorded lateral and longitudinal acceleration of the vehicle. It highlights the combined slip situation at which the vehicle is driven. Fig. \ref{fig:Hock_DD} shows the level of distrust in the NN for the hybrid approaches.}
    \label{fig:Hock}
\end{figure}

Similar conclusions are obtained from the spiral manoeuvre represented in Fig. \ref{fig:spiral}. Here, the CNN-UKF approach outperforms the accuracy of all other three approaches, particularly from \SI{5}{s} to \SI{13}{s}. The performance of the CNN-UKF is similar to sum of the best estimation between the data-driven approach, from \SI{4}{s} to \SI{6}{s}, and the model-based approach, from \SI{6}{s} to \SI{10}{s}. \compressParag

The test set also contains a recording of an entire lap in a racing circuit, where the effect of combined slip is maximal. Fig. \ref{fig:Hock_Acc} shows the vehicle's lateral and longitudinal acceleration, and it highlights how the driver is pushing the vehicle at the limit of handling in all the corners, see $\left[1,\, 7\right]$ \SI{}{s}, $\left[16,\, 19\right]$ \SI{}{s} and $\left[23,\, 29\right]$ \SI{}{s}. The sideslip angle estimation performance of the four approaches is represented in Fig. \ref{fig:Hock_Sideslip}. The model-based approach has the lowest performance, especially in the range $\left[1,\, 7\right]$ \SI{}{s}. This result is expected because the implemented Dugoff tyre model works in pure slip conditions. A similar conclusion is also visible in the spiral manoeuvre, Fig. \ref{fig:spiral}. The data-driven approach performs better than a model-based approach. However, it has the maximum absolute error at \SI{23}{s} and \SI{26}{s}, where the vehicle performs a cornering while braking. Both proposed hybrid approaches have higher accuracy than the others, but the CNN-UKF has the best performance. A possible explanation is the physic-informed NN architecture, which allows evaluating a very accurate NN level of distrust. Fig. \ref{fig:Hock_DD} shows the NN level of distrust for the DE-UKF and CNN-UKF. While the DE-UKF level of distrust is almost constant along the manoeuvre, the one associated with CNN-UKF has two peaks in correspondence with the data-driven maximum errors. This allows the CNN-UKF to avoid following the high estimation error of the data-driven component. It is further proof of how the CNN-UKF is a physics-informed NN in which the UKF and NN are mutually cooperating to improve the overall estimation of the hybrid approach.

\subsection{Robustness Analysis using the Limited Dataset}
\label{subsres2}
\noindent A sideslip angle filter must not only be accurate, but it should be robust to a different amount of qualitative data during the training and tuning phase. Hence, to prove the robustness of the proposed hybrid approach, its estimation performance is compared with the baseline methods when they all have been trained using the limited dataset.
\begin{table}[htp]
    \caption{Sideslip angle estimation comparison using the limited dataset.}
    \label{tab:Sub_Data_Comp}
    \begin{center}
    \begin{tabular}{ >{\centering\arraybackslash}m{1in}  >{\centering\arraybackslash}m{0.3in}  >{\centering\arraybackslash}m{0.3in} >{\centering\arraybackslash}m{0.3in}  >{\centering\arraybackslash}m{0.3in} }
    \toprule[1pt]    
    \textbf{Approaches} &  \textbf{MSE [\SI{}{deg^2}]}  &  \textbf{MSE\textsubscript{nl} [\SI{}{deg^2}]} &  \textbf{ME [\SI{}{deg}]} &  \textbf{ME\textsubscript{nl} [\SI{}{deg}]}\\
    \hline    
    Model-based         & 0.161             & 0.277             & 1.111             & 0.991 \\
    Data-driven         & 0.223             & 0.358             & 1.445             & 1.284 \\
    DE-UKF              & 0.157             & 0.270             & 1.103             & 0.983\\
    \textbf{CNN-UKF}    & \textbf{0.156}    & \textbf{0.269}    & \textbf{1.099}    & \textbf{0.975} \\
    \bottomrule[1pt]
    \end{tabular}
    \end{center}
\end{table}

The overall comparison is presented in Table \ref{tab:Sub_Data_Comp}. Here, the model-based approach shows the same performance as with the full dataset (see Table \ref{tab:Optim_Data_Comp}), because it is not influenced by the amount of data. As expected the other approaches show a reduced performance with the limited dataset where the MSE is more than doubled while the ME sees a moderate increase. Now the data-driven approach has the worst performance in all four KPIs. The accuracy loss is higher than \SI{30}{\%} for all the indicators, without a particular weakness in one of the proposed KPIs. An explanation is that the dataset does not have representative data of the vehicle driven with $|a_y|\geq$ \SI{7}{m/s^2}, so it must generalise much more than with a full dataset. Significantly, the NN must reconstruct the extreme non-linear vehicle behaviour, the most complex vehicle operating region, without having representative data for these conditions.

The model-based and hybrid approaches' performance is very similar, but DE-UKF and CNN-UKF have the best KPIs. The explanation is that hybrid approaches use the best estimation accuracy of the model and the NN together. Both hybrid approaches strongly rely on the estimation of the UKF process model because they cannot put much trust in the data-driven part. However, the NN still has benefits when the vehicle behaves linearly due to the excellent amount of data in that range. This highlights how the hybrid approaches improve the robustness of both model-based and data-driven approaches. The hybrid approach shows a minor improvement compared with the model-based approach. However, the result is significant because it highlights how the hybrid approach is as robust as a model, even if trained with a limited training dataset. On the contrary, a purely data-driven approach is not robust for using a small training set resulting in poor estimation accuracy.\compressParag

The CNN-UKF performs slightly better than DE-UKF in all four KPIs. However, the CNN-UKF outperforms the DE-UKF, mainly for the MSE and MSE\textsubscript{nl}. The main reason is the adaptability of the process noise parameters, which cope with the change of vehicle model mismatches in the various vehicle operating points. However, the improvement in accuracy is not enough to be considered significant ($< 5$ \%). 
\begin{figure}[!t]
    \centering
    \includegraphics[height=3.5cm, keepaspectratio]{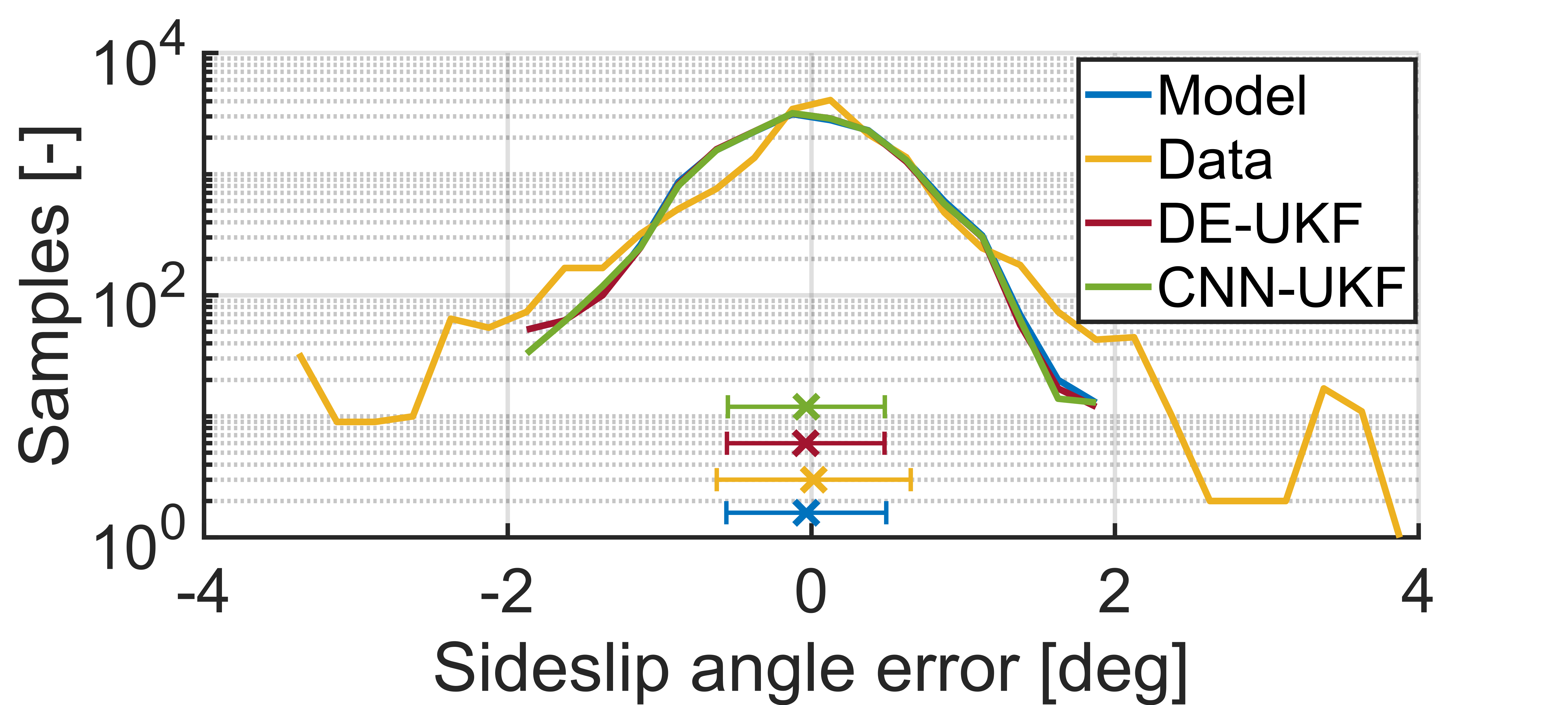}
    \caption{Distribution of the sideslip angle error when the vehicle $|a_y|>$ \SI{4}{m/s^2} for every approach in the test set. Each bin is \SI{0.25}{deg} wide. The x represents the mean and the line between the vertical symbols ($|-|$) is the standard deviation of the sideslip angle error. Results based on the limited dataset, see Fig. \ref{fig:J_turn_Sub} the for best results. }
    \label{fig:Dist_Opt_Non_Sub}
\end{figure}
\begin{figure}[!t]
    \centering
    \includegraphics[height=3.5cm, keepaspectratio]{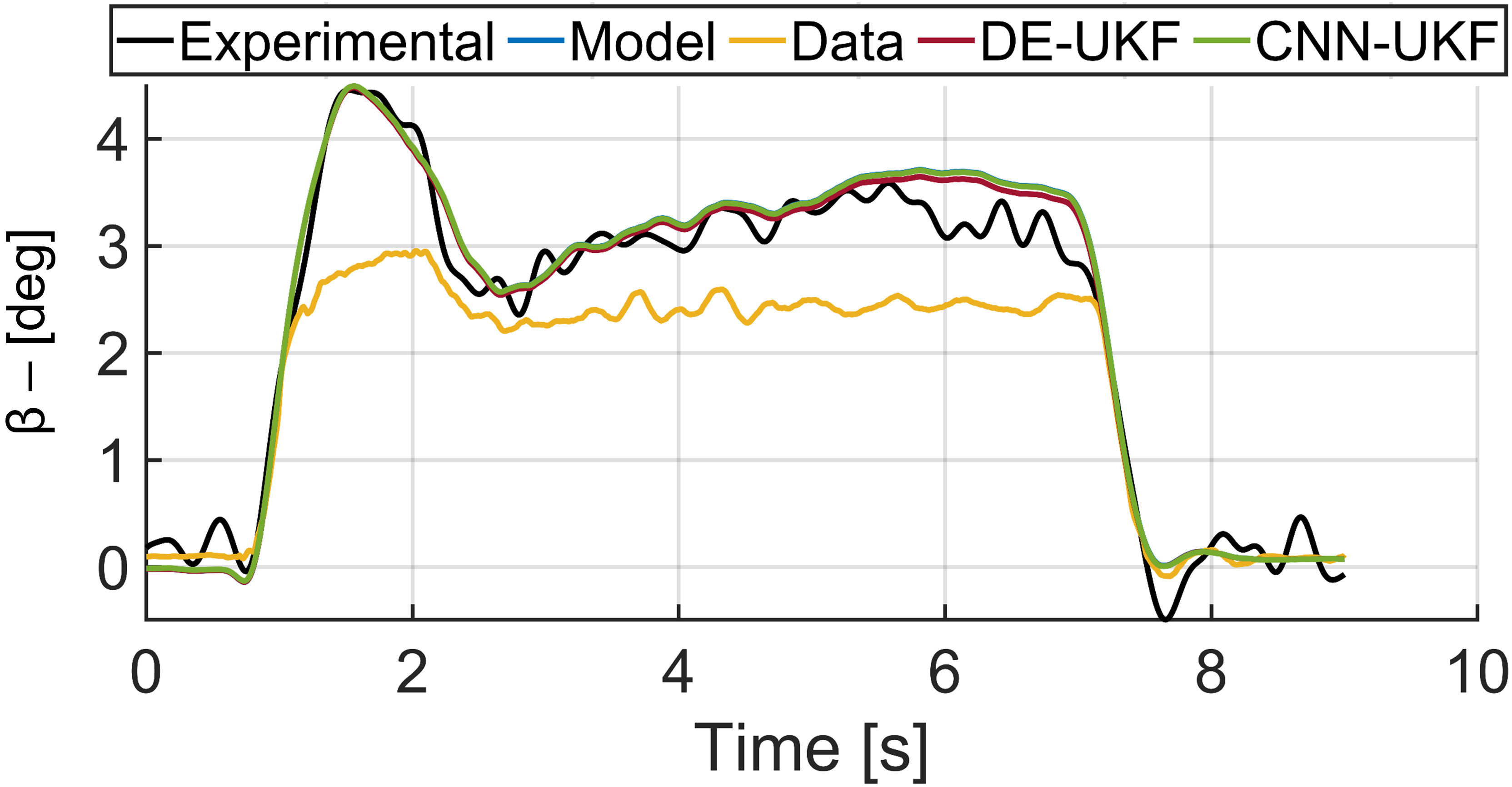}
    \caption{J-turn manoeuvre. Comparison of the sideslip angle estimation between all four approaches using the limited dataset.}
    \label{fig:J_turn_Sub}
\end{figure}
\begin{figure}[!t]
    \centering
    \includegraphics[height=3.5cm, keepaspectratio]{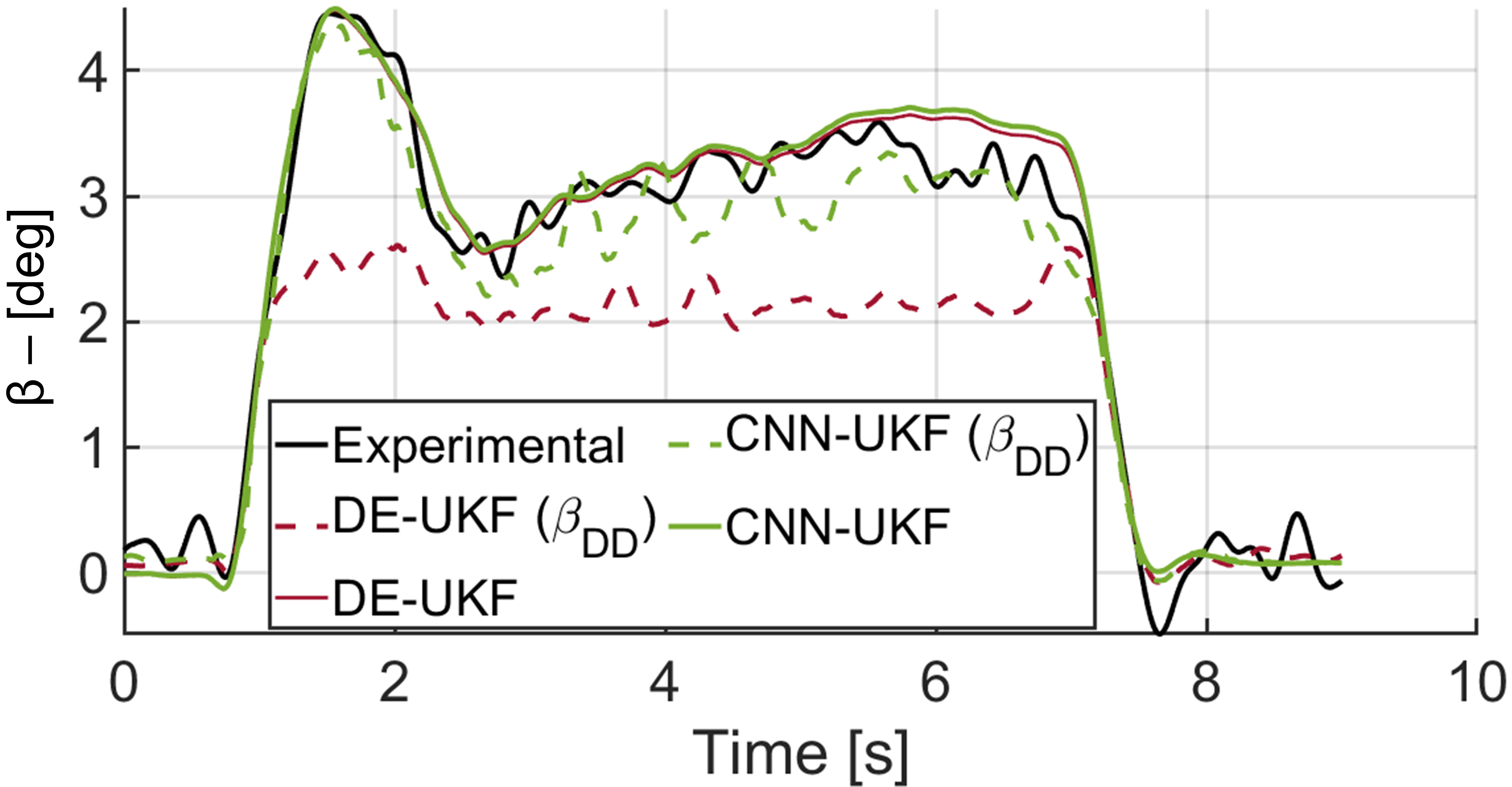}
    \caption{J-turn manoeuvre. Comparison of the estimated $\beta$ and $\beta_{DD}$ between the hybrid approaches using the limited dataset.}
    \label{fig:J_turn_Sub_DD}
\end{figure}

Fig. \ref{fig:Dist_Opt_Non_Sub} shows the sideslip angle error log distribution in the non-linear operating range. All the approaches which rely on a model highly outperform the data-driven approach. The latter have $\beta$ error samples in the range $\left[-3.8,\,4\right]$ deg, while the other approaches have $\beta$ errors between $\left[-1.8,\ 1.8\right]$ deg. This proves that the data-driven approach is highly prone to high estimation errors when trained with a limited dataset. The performance of the model-based and hybrid approaches is very similar. They also share an equal error distribution. The data-driven approach slightly outperforms the other approaches in the very low error range  $\left[-0.3,\,0.6\right]$ deg. This explains why the hybrid approaches are more accurate overall than the model-based one, despite mainly relying on it.

Fig. \ref{fig:J_turn_Sub} shows the sideslip angle estimation in a J-turn manoeuvre at the handling limits. The model-based and hybrid approaches behave almost identically, and all strongly outperform the purely data-driven approach. The only visible differences are between $\left[1.5,\,3\right]$ s, where the CNN-UKF captures slightly better the conclusion of the peak and between $\left[4,\,7\right]$ s where the DE-UKF is closer to the $\beta$ reference.

The major difference between DE-UKF and CNN-UKF is visible from the comparison of the $\beta_{DD}$, see Fig. \ref{fig:J_turn_Sub_DD}. The $\beta_{DD}$ computed by the CNN-UKF is highly outperforming the one estimated by the DE-UKF. The explanation is that the CNN-UKF is trained end-to-end, so the output of the CNN has physical information that the NN uses to increase its accuracy. The DE is trained independently, performing similarly to the purely data-driven approach. An higher accuracy of $\beta_{DD}$ implies that the following UKF can rely on a better sideslip angle pseudo-measurement. This proves the benefits of using a physical informed-NN. Despite this, the $\beta$ estimation of DE-UKF and CNN-UKF is similar because the model-based approach still outperforms both $\beta_{DD}$. Thus, both hybrid approaches mainly rely on the UKF. Due to the high chances of dealing with a limited dataset, the hybrid approach is fundamental to improving vehicle sideslip angle estimation.

However, the performance of the proposed CNN-UKF approach is still influenced by the amount and quality of data in the training set. Thus, it still represents a limitation of the proposed approach that must be addressed in the future. This highlights the importance of defining standards procedure to collect valuable and broad datasets. Regardless, the proposed CNN-UKF allows the introduction of possible solutions for lack of data, e.g., weakly-supervised learning during the end-to-end training, which allows for using data recorded without expensive sensors. \compressParag

\section{Conclusion}
\label{conc}
\noindent The paper presents a novel hybrid approach to vehicle sideslip angle estimation, which involves utilising the physical knowledge from a UKF based on a single-track vehicle model to enhance the estimation accuracy of a CNN. Using a large-scale experimental dataset of 216 manoeuvres, it has been shown that the hybrid approach is more accurate than purely model-based or data-driven approaches. Moreover, the CNN-UKF is slightly reducing the MSE of the DE-UKF. However, when the MSE\textsubscript{nl} is compared, the CNN-UKF outperforms the DE-UKF by \SI{25}{\%}, providing a much higher accuracy in the most critical operating region for active vehicle control systems. The CNN-UKF, thanks to the end-to-end training, is forcing the CNN to comply with the vehicle physics, reducing the ME and ME\textsubscript{nl} of all other approaches. When a limited dataset is provided, the proposed hybrid approach has a minor improvement in the estimation robustness over the model-based and the DE-UKF approach for all the KPIs. The CNN-UKF is highly outperforming the estimation of a purely data-driven approach. Future works involve testing the generalisation capability of the CNN-UKF utilising a dataset with different levels of road grip, e.g. wet, snow or icy roads.
\bibliographystyle{IEEEtran}
\bibliography{IEEEabrv, references.bib}

\begin{IEEEbiography}[{\includegraphics[width=1in,height=1.25in,clip,keepaspectratio]{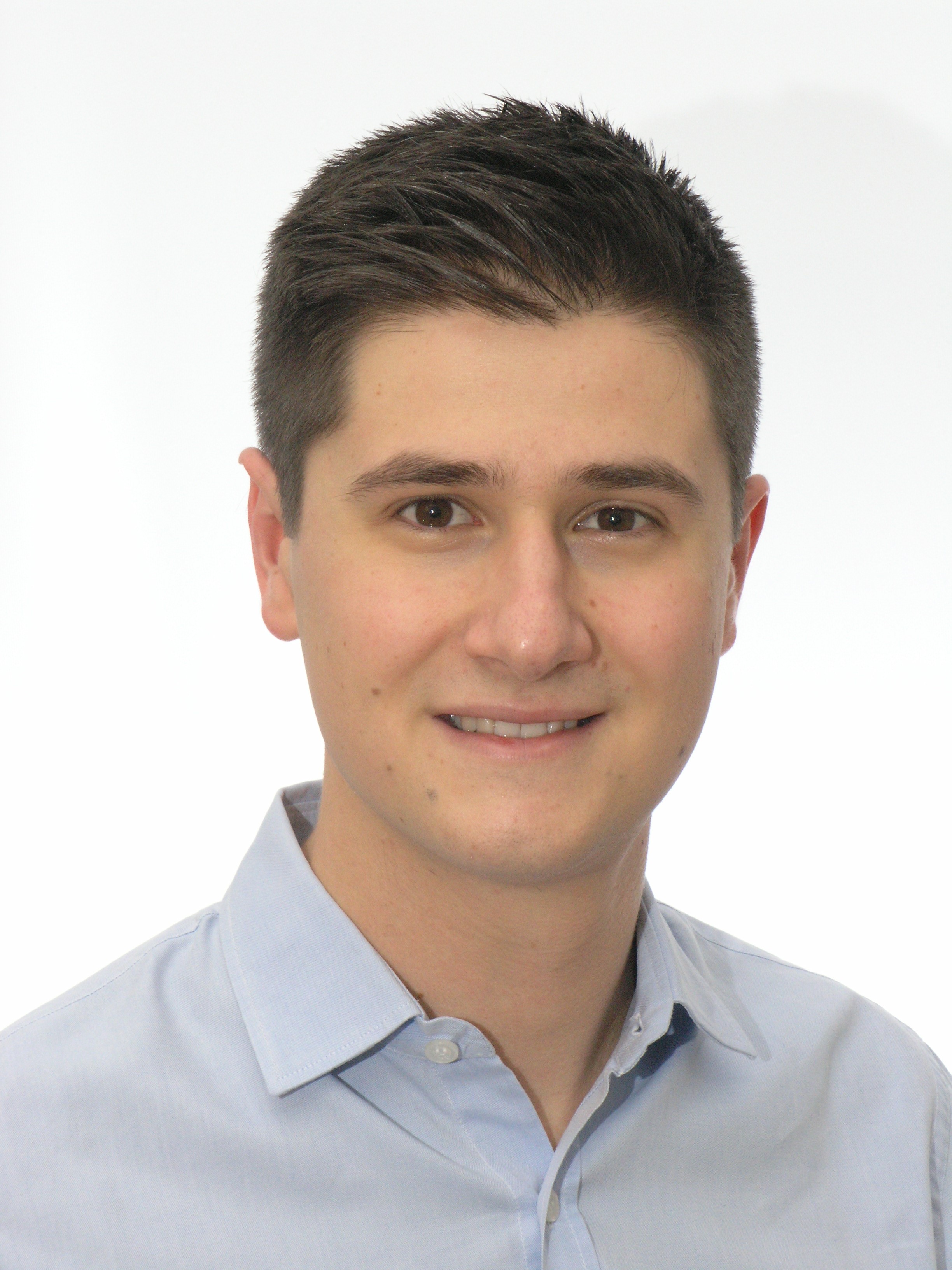}}]{Alberto Bertipaglia} received the  M.Sc. degree, (cum laude), in automotive engineering from the Politecnico di Torino, Italia, in 2020. He is pursuing the Ph.D. degree in the Department of Cognitive Robotics, at the Delft University of Technology. His work is part of the national research project, aiming to develop innovative active vehicle controls to improve vehicle’s stability and increase the safety of automated vehicles performing evasive manoeuvres.
\end{IEEEbiography}

\begin{IEEEbiography}[{\includegraphics[width=1in,height=1.25in,clip,keepaspectratio]{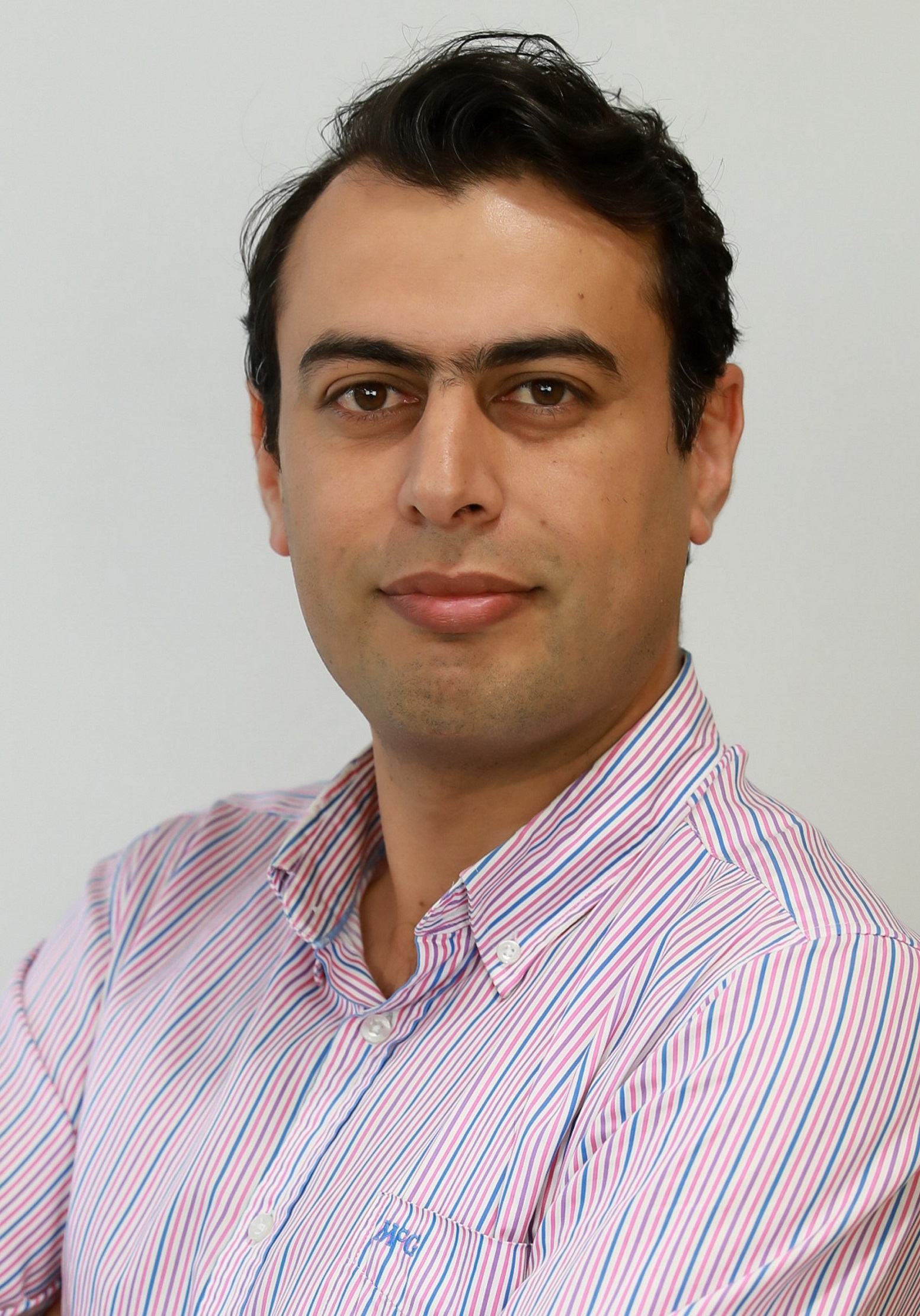}}]{Mohsen Alirezaei} received the PhD degree in Mechanical Engineering, Robotics and Control in 2011 and was a postdoc researcher at Delft University of Technology in 2012. He was a Senior Scientist with Integrated Vehicle Safety Department, TNO automotive during 2012–2019 and a part time Assistant Professor with the Delft University of Technology during 2015–2019. He is currently a Fellow Scientist with Siemens Industry Software and Services, Helmond, The Netherlands and a part time Assistant Professor with the Eindhoven University of Technology, Eindhoven, The Netherlands. His research interests include verification and validation of automated and cooperative automated driving and advance driver assistance systems.
\end{IEEEbiography}

\begin{IEEEbiography}[{\includegraphics[width=1in,height=1.25in,clip,keepaspectratio]{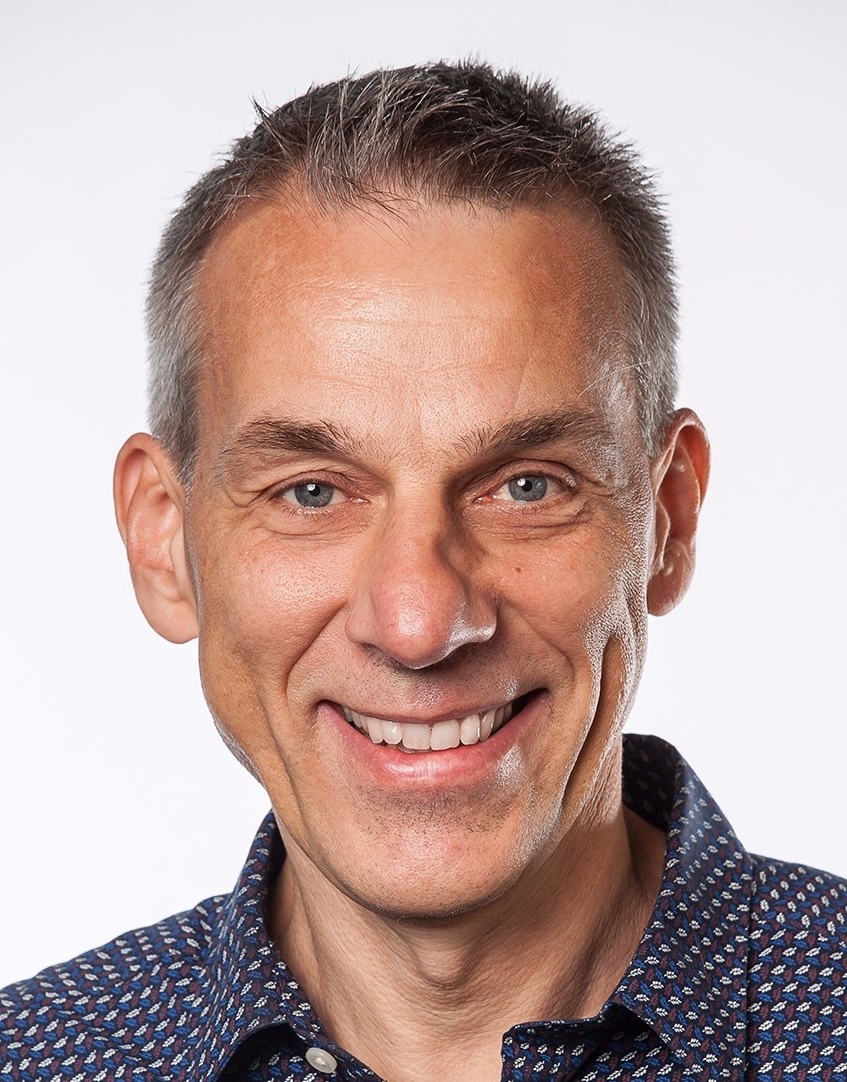}}]{Riender Happee} received the Ph.D. degree from TU Delft, The Netherlands, 1992. He investigated road safety and introduced biomechanical human models for impact and comfort at TNO Automotive (1992-2007). Currently, he investigates the human interaction with automated vehicles focusing on safety, motion comfort and acceptance at the Delft University of Technology, the Netherlands, where he is a Full Professor.
\end{IEEEbiography}

\begin{IEEEbiography}[{\includegraphics[width=1in,height=1.25in,clip,keepaspectratio]{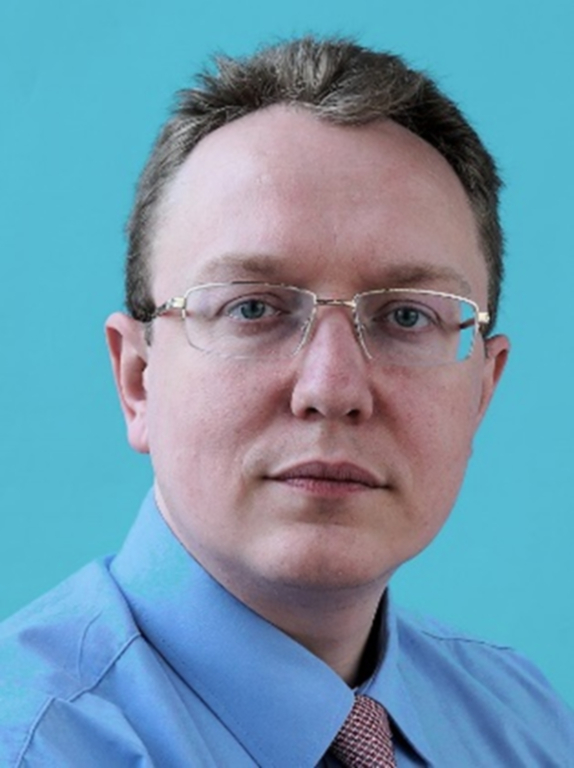}}]%
{Barys Shyrokau} received the DiplEng degree (cum laude), 2004, in Mechanical Engineering from the Belarusian National Technical University, and the joint PhD degree, 2015, in Control Engineering from Nanyang Technological University and Technical University Munich. He is an assistant professor in the Section of Intelligent Vehicles, at the Delft University of Technology, and his research interests are vehicle dynamics and control, motion comfort, and driving simulator technology. Scholarship and award holder of SAE, FISITA, DAAD, SINGA, ISTVS, and CADLM. \compressParag
\end{IEEEbiography}

\end{document}